\documentstyle[12pt,epsf]{article}
\def\beq{\begin{equation}}
\def\eeq{\end{equation}}
\def\bea{\begin{eqnarray}}
\def\eea{\end{eqnarray}}
\def\R{\rangle}
\def\L{\langle}
\def\lt{\left}
\def\rt{\right}
\newcommand{\sectiono}[1]{\section{#1}\setcounter{equation}{0}}
\newcommand{\subsectiono}[1]{\subsection{#1}}

\begin{document}

\begin{flushright}
hep-th/0011047\\
MRI-P-001101
\end{flushright}

\vskip 3.5cm

\begin{center}
{\large \bf Unstable Non-BPS D-Branes of Type-II String Theories

\medskip

in Light-Cone Green-Schwarz Formalism}\\

\vspace*{6.0ex}

{\large \rm Partha Mukhopadhyay\footnote{\tt partha@mri.ernet.in}} 

\vspace*{1.5ex}

{\large \it Harish-Chandra Research Institute\footnote{Formerly 
Mehta Research Institute of Mathematics \&\ Mathematical
 Physics}}\\  
{\large \it  Chhatnag Road, Jhusi, Allahabad 211019, India}

\vspace*{4.5ex}

{\bf Abstract}

\begin{quote}

The problem of describing the boundary states of unstable non-BPS D-branes of 
type-II string theories in light-cone Green-Schwarz (GS) formalism is 
addressed. Regarding the type II theories in light-cone gauge as different 
realizations of the $\hat{SO}(8)_{k=1}$ Kac-Moody algebra, the non-BPS D-brane 
boundary states of these theories are given in terms of the relevant Ishibashi 
states constructed in this current algebra. Using the expressions for the 
current modes in terms of the GS variables it is straightforward to reexress 
the boundary states in the GS formalism. The problem that remains is the lack 
of manifest $SO(8)$ covariance in these expressions. We also derive the various
 known expressions for the BPS and non-BPS D-brane boundary states by starting 
with the current algebra Ishibashi states. 

\end{quote}
\end{center}

\newpage

\tableofcontents
\sectiono{Introduction and Summary}
\label{sec-int}
The ten dimensional type IIA and IIB superstring theories admit BPS $Dp$
-branes \cite{db, divecchia} for even and odd $p$ respectively. The restriction
  on $p$ 
comes from the condition that half of the supersymmetries (SUSY) of the bulk 
be preserved in the world-volume theory. It has been realized recently that IIA
 and IIB theories possess D-branes of respectively odd and even dimensions 
also, which are called non-BPS D-branes \cite{bg, asen1, asen2} (for reviews 
and complete list of references see \cite{asen3, gaberdiel}), as they break all
 the SUSY of the bulk. As it is well known that string theory has two types of
formalisms, namely NSR (Neveu-Schwarz-Ramond)\cite{nsr} and GS
(Green-Schwarz)\cite{gsw} formalisms, it is desirable to have descriptions of 
all these nonperturbative objects in both the formalisms. This is because, in 
different situations different formalisms become useful. In most of the 
situations (at least in the flat space), NSR formalism has been useful where we
 have proper description of all the D-branes in terms of the NSR variables 
(see \cite{db, divecchia, bg, asen1, asen2, asen3, gaberdiel, nsr, bstate1, 
bstate2}). One specific 
use of the GS description would be to check SUSY of a D-brane. Since the SUSY 
generators\cite{gsw, gg} are simple in terms of the GS fields, doing this would
 be easier if the D-brane boundary state\cite{bstate1} (for D-brane boundary 
states in light-cone gauge see \cite{bg, gaberdiel, bstate2, g, gg}) is written
 in terms of the GS variables. The boundary states for the BPS D-branes in GS 
formalism have already been found by Green\cite{g} and Green and 
Gutperle\cite{gg}. In this 
paper we address the problem of describing the unstable non-BPS D-branes  
in ten flat dimensions in light-cone GS formalism \cite{gsw}. More specifically
 we will concentrate only on the fermionic part of the world-sheet theory as 
the bosonic part is the same in the two formalisms. Throughout this paper the 
phrase ``non-BPS branes'' would imply the unstable non-BPS D-branes of IIA and 
IIB string theories in ten dimensional Minkowski space.

Let us first discuss the problem without going into the technical details. A 
D-brane has two different descriptions --- open string world-volume 
description and 
closed string boundary state description. In the first approach, one puts 
proper open string boundary conditions on the fields living on the fundamental
string world-sheet. The tangential and 
transverse directions to the brane have respectively Neumann and Dirichlet 
boundary conditions. A space-time parity 
transformation along a single direction only on the right moving 
(antiholomorphic) variables interchanges the Neumann and Dirichlet boundary 
conditions along that direction \cite{boson, nsr}. Starting from any BPS 
boundary condition in a given 
theory one generates all the non-BPS boundary conditions by applying all 
possible parity transformations involving only odd number of directions. Since 
the world-sheet fermions in NSR formalism form a vector representation of the 
space-time group $SO(9,1)$\footnote{In light-cone gauge this comes down to 
$SO(8)$.}, they transform in a simple way under parity. Hence in these 
variables the boundary 
conditions for non-BPS D-branes are as simple as those for BPS branes, and the 
conformal field theory on the world-sheet has a simple description. But the 
world-sheet fermions in the light-cone
GS formalism form spinor representations of $SO(8)$. A parity transformation 
along an odd number of directions converts a spinor representation to a 
conjugate spinor representation \cite{liealg} which does not belong to the set 
of variables used for defining the theory. This complication makes 
it difficult to describe the boundary conformal field theory for a non-BPS 
D-brane in GS 
formalism. Because of the same difficulty, the usual procedure of converting 
the open string boundary conditions to the closed string gluing conditions and 
then solving that in the form of a boundary state, does not work. 

One way to solve the problem may seem to translate the known non-BPS
boundary state in NSR formalism in terms of the GS variables. But the field 
redefinition required to go from the NSR variables to the GS variables is too
complicated to do this. Thus we need to follow a different approach. For this
let us try to understand the problem described, in a 
more fundamental way. The reason that we have different formalisms like NSR and
 GS of the same theory (type IIA or IIB) \cite{gsw, witten} is the existence of
 an underlying local current algebra\cite{go, fuchs}, namely  
$\hat{SO}(8)_{k=1}$ and the crucial property of the group $SO(8)$ that it has 
three eight dimensional representations, namely the vector, spinor and 
conjugate spinor representations\footnote{See ref.\cite{gsw} for a discussion
on `triality', which is a symmetry of the $SO(8)$ Dynkin diagram under which 
these three fundamental and inequivalent representations tranform among each 
other.}\cite{liealg}. This makes it possible to 
formulate a realization of the $\hat{SO}(8)_{k=1}$ current algebra in different
 languages\cite{go}
 like that of NSR and GS, where in the first case the world-sheet 
fermions form the vector representation and in the second case spinor and/or
conjugate spinor representation. From this point of view the problem at hand is
 that of expressing the boundary state of a non-BPS D-brane in a specific 
realization (GS)  of the current algebra $\hat{SO}(8)_{k=1}$. Our strategy will
 be to analyse the D-brane boundary states in these theories in terms of the 
current algebra itself \cite{Ishib, ko}. Although going from one formalism to 
another is in general complicated, expressing the currents in any formalism is 
 straightforward. Thus once we get a boundary state in terms the local 
$SO(8)$ currents, that can be reexpressed in the GS formalism.

We proceed along the following steps:
\begin{enumerate}
\item We consider $SO(8)$ Kac-Moody current algebra of level $k=1$ associated 
with the NSR/GS fermions and its irreducible unitary highest weight 
representations (irreps) \cite{go}. The type IIA and IIB theories are regarded
as different realizations of the same current algebra. These two theories 
differ by the irreps that are realized.
\item 
Given a consistent boundary condition on the currents, one can construct a 
unique 
Ishibashi state \cite{Ishib, ko} for each irrep. Therefore given a boundary 
condition 
relevant for a specific D-brane, there is an Ishibashi state for each irrep of
 $\hat{SO}(8)_{k=1}$. Not all the Ishibashi states are important for the
 present purpose. Only those, the corresponding irreps for which are realized 
in a given theory, are needed to construct the boundary states for D-branes of 
that theory. It turns out that if the boundary condition is BPS in a given 
theory then there are two realizable Ishibashi states and if it is non-BPS 
then there is only one. In the first case the two Ishibashi states are the 
NS-NS and the R-R parts of the corresponding D-brane boundary state and in the 
second case the Ishibashi state itself is the non-BPS boundary state\footnote{
Actually the Ishibashi states considered themselves do not construct the 
complete D-brane boundary states, but rather form only the fermionic oscillator
 part of that. We will make it more precise in subsec.~\ref{subsec-IIBbranes}.
}. 
\item
To write down an Ishibashi state explicitly one needs to consider a complete 
set of basis vectors for the Hilbert space \cite{ko}. One can construct various
 sets of basis states. It can be constructed in a given formalism by applying 
the corresponding negative mode fermionic oscillators on the ground 
state(s). This is the natural fermionic oscillator basis in that formalism. The
basis can also be 
constructed purely in terms of the current modes, which is independent of any 
formalism. We will show that in a given formalism, choosing the fermionic
 oscillator basis to work with, directly reproduces the standard form of the 
already known boundary states. This means that the boundary 
states for both the BPS and non-BPS D-branes in the NSR formalism and only the
BPS D-branes in the GS formalism can be obtained in this way. We will see that 
in the current algebraic language the present problem can be stated by saying 
that the natural fermionic oscillator basis arising in the GS formalism is not 
suitable for writing down the Ishibashi states corresponding to the non-BPS 
D-branes\footnote{This is related to the fact that implementation of the 
non-BPS boundary conditions on the spinor variables is not as easy as it is in 
the case of variables transforming as vectors.}.
\item
We will argue that it is preferable to construct the basis in terms of the 
current modes for writing down the Ishibashi states corresponding to 
the non-BPS D-branes. Since the currents transform as tensors under parity    
tranformation, implementation of both the BPS and non-BPS boundary conditions 
is quite straightforward. But there is a principal drawback of this type of 
basis. The free module constructed out of the current modes contains null 
states as opposed to the case of various fermionic oscillator basis. To use it 
to construct the Ishibashi states one needs to make the basis free of null 
states. This can be done, but one loses explicit covariance during this 
process. Given this basis, one can reexpress this in terms of the Green-Schwarz
 variables, but the resulting expression is not manifestly $SO(8)$ covariant. 
The lack of manifest covariance is the principal drawback of this approach.
\end{enumerate}

\noindent {\bf Organization of the paper:} In Sec.~\ref{sec-curnt.alg} we 
briefly discuss the affine $\hat{SO}(8)_{k=1}$ current algebra both in the 
covariant and the Cartan-Weyl basis and quote the basic results 
about its irreps which will be needed for the present purpose. In 
Sec.~\ref{sec-realization} we discuss how $\hat{SO}(8)_{k=1}$ is realized in 
NSR, GS and bosonic formalism. Here the `bosonic formalism' 
considered\footnote{It is not true that in principle one needs to cosider the 
bosonic formalism in the present approach. We need to consider it because of 
the following reason: the non-covariant basis that we will construct in terms 
of the current modes 
will have exact resemblance (quantum number wise) with the natural basis that 
one would construct in the bosonic formalism.} is the 
one which is obtained by performing the abelian bosonization\footnote{
Throughout this paper we will use the word `bosons' to refer only the abelian
bosons. The non-abelian bosons will always be called 
`currents'.}\cite{bosonization} of the NSR or
 the GS fermions suitably on the world-sheet. These sections establish the
 notations and the basic definitions for the various variables that we will 
work with. In Sec.~\ref{sec-Ishib} we introduce the 
current algebra Ishibashi 
states, and demonstrate how using the present current algebraic approach one 
can derive the known results for BPS and non-BPS D-brane boundary states by 
constructing the basis states in terms of the appropriate fermionic 
oscillators. At this point, we indicate the problem of using the similar basis
for constructing the boundary states for non-BPS D-branes in the GS formalism.
In Sec.~\ref{sec-basis} we give the explicit construction of an orthonormal
basis in terms of the current modes and then show how to get the non-BPS 
D-brane boundary 
states in GS formalism by using this basis. The appendices contain some 
additional technical details. Appendix \ref{app-vac.rep} contains the explicit 
construction of the current algebraic vacuum representations on which the 
irreps are built up by applying the current modes. The specific representation 
chosen for the $SO(8)$ gamma matrices is given in appendix \ref{app-gamma}. 
Appendix \ref{app-basis} shows that the bosonic vacuum states 
defined in Sec.~\ref{sec-basis} are normalised to $1$, which supports the 
proof (given in Sec.~\ref{sec-basis}) of orthonormality of the constructed 
basis.

\sectiono{The Affine $\hat {SO} (8)_{k=1}$ Current Algebra and its Irreducible
Unitary Highest Weight Representations}
\label{sec-curnt.alg}
The affine Kac-Moody algebras, their representations, realizations in terms of 
fermionic and bosonic fields and specifically superstring theory in the context
of $\hat{SO}(8)_{k=1}$ have been discussed quite extensively and in a 
self-contained manner in ref.\cite{go}. The purpose of this section is to
 introduce notations, basic definitions and review results only for 
$\hat {SO} (8)_{k=1}$, which we will need for the present paper. 

The infinite dimensional Lie algebra $\hat {SO} (8)_{k=1}$ is given by:
\bea
[J^{\mu \nu}_m, J^{\rho \sigma}_n ] &=& i( \delta^{\nu \rho} J^{\mu \sigma}_
{(m+n)} + \delta^{\mu \sigma} J^{\nu \rho}_{(m+n)} - \delta^{\mu \rho} 
J^{\nu \sigma}_{(m+n)} - \delta^{\nu \sigma} J^{\mu \rho}_{(m+n)} )\nonumber \\
&& + m \delta_{(m+n),0} ( \delta^{\mu \rho}  \delta^{\nu \sigma} - 
\delta^{\mu \sigma}\delta^{\nu \rho} )~,  \label{JJ}
\eea
where $J^{\mu \nu}_n= - J^{\nu \mu}_n$ $(\mu ,\nu =1,2,...,8, n\in {\bf Z})$ are the 
elements of the algebra. We define the Cartan-Weyl (CW) basis as follows:
\beq
H^{i}_n = J_n^{2i-1,2i},\quad \quad \quad (i=1,..,4)~, \label{HJ}
\eeq
\beq
E_n^{\alpha} = \frac{1}{2} [(J_n^{2i-1,2j-1} - \eta_{1} \eta_{2} J_n^{2i,2j})
-i(\eta_{1} J_n^{2i,2j-1} + \eta_{2} J_n^{2i-1,2j} )]_{i<j}~,  \label{EJ}
\eeq
where $\alpha = (\eta_1 e_i + \eta_2 e_j)_{i<j} $ is a nonzero root, $\eta_1,
\eta_2 = \pm 1 $ and $e_i$'s are the  orthonormal basis of the Cartan 
subalgebra (CSA) space:
\beq
e_1 = \pmatrix{1\cr0\cr0\cr0}~,\quad  e_2 = \pmatrix{0\cr1\cr0\cr0}~, \quad
e_3 = \pmatrix{0\cr0\cr1\cr0}~,\quad e_4 = \pmatrix{0\cr0\cr0\cr1}~. \label{e}
\eeq
In the CW basis the algebra-elements are the twenty-four $E^{\alpha}_n$'s 
and four $H^i_n$'s ($\forall n \in {\bf Z}$), in terms of which the algebra 
takes the following form:
\bea
\lt[ H_m^i,H_n^j \rt] &=& m \delta^{ij} \delta_{m+n,0}       ~,\label{HH} \\
\lt[ H_m^i,E_n^{\alpha} \rt] &=& (\alpha .e_i) E_{(m+n)}^{\alpha} ~,\label{HE}
 \\
\lt[ E_m^{\alpha}, E_n^{\beta} \rt] &=& \lt\{
\begin{array}{ll}
\epsilon(\alpha,\beta)E_{(m+n)}^{\alpha + \beta}\quad
&\mbox{if } (\alpha + \beta)\mbox{ is a root}~,\\
{}&\mbox{{\it i.e.\ }} \alpha .\beta = -1~,\\
{}&{}\\
(\alpha .e_i) H_{(m+n)}^i + m \delta_{m+n,0}\quad
&\mbox{if } \alpha = -\beta ~,\\
{}&\mbox{ {\it i.e.\ }}\alpha .\beta = -2 ~,\\  
{}&{}\\
0&\mbox{otherwise}~,
\end{array}\rt.
 \label{EE}
\eea
Here $\epsilon(\alpha, \beta)$ is a phase antisymmetric in $\alpha$ and
$\beta$ when $\alpha .\beta = -1$ and can be computed explicitly using eqns.
(\ref{JJ}) and (\ref{EJ}). The zero-modes elements satisfy the finite 
dimensional $SO(8)$ algebra. Out of the twenty-four non-zero roots the 
following are the twelve positive roots: 
\bea
\alpha_1 =\psi =e_1 + e_2~, & \alpha_2 =\alpha_1^s =e_1 - e_2~, & 
\alpha_3 =\alpha^s_4 =e_3 + e_4~, \nonumber\\
\alpha_4 =\alpha^s_3 =e_3 - e_4~, & \alpha_5 =e_1 + e_3~, & 
\alpha_6 =e_1 - e_3~, \nonumber\\
\alpha_7 =e_1 + e_4~, & \alpha_8 =e_1 - e_4~, & \alpha_9 =e_2 + e_3~, 
\nonumber\\
\alpha_{10} =\alpha^s_2 =e_2 - e_3~, & \alpha_{11} =e_2 + e_4~, & 
\alpha_{12} =e_2 - e_4.    \label{roots}
\eea                         
where $\psi$ is the highest root and $\alpha^s_i$ $(i=1,...,4)$ are the simple 
roots. Now we define various conformal fields (local currents) on the $z$-plane
 with the algebra elements as their modes:
\bea
{\cal J}(z) &=& \sum_{n\in {\bf Z}} \frac{{\cal J}_n}{z^{n+1}}
~,\label{JHEcurrents}
\eea
where ${\cal J}$ stands for any of $J^{\mu \nu}$, $H^i$ and $E^{\alpha}$.
We connect the $z$-plane with the closed string tree world-sheet (cylinder) of 
the type II theories by the conformal map $z = e^{t} e^{-i\sigma}$, where 
$(t,\sigma)$ parameterize the cylinder with the range: $-\infty < t < 
\infty$ , $0 \leq \sigma < 2\pi$. Sugawara's construction for the 
energy-momentum tensor reads\cite{go, fuchs, nsr}:
\bea
T^{SUG}(z) &=& \frac{1}{14}:JJ:(z)~,
\cr
&=& \sum_{n\in {\bf Z}} \frac{L^{SUG}_n}{z^{n+2}} ~,\label{Tsug} 
\eea
where, the factor of $14$ comes from $2k + C_2$, $C_2$ being the quadratic 
Casimir in the adjoint representation. For $SO(8)$, $C_2=12$. The composite
object $:JJ:(z)$ has the following definition\cite{nsr}:
\bea
:JJ:(z) = \lim_{w\rightarrow z} \lt( \sum_{\mu < \nu} J^{\mu \nu}(z) 
J^{\mu \nu}(w) - \frac{28}{(z-w)^2} \rt)~.  \label{JJ.conf.ord}
\eea
The modes $L^{SUG}_n$ satisfy a Virasoro algebra with central charge $c=4$. 
\\

\noindent{{\bf Irreducible Unitary Highest Weight Representations:}}

\noindent Here we will discuss the results of the unitary highest weight 
representations
which are irreducible under the Kac-Moody algebra $(\ref{JJ})$. A highest 
weight state $|w\R$ for a given weight vector $w=\sum_i \lt(w\rt)_i e_i$ is 
defined in the following way:
\bea
\begin{array}{rll}
E^{\alpha}_0 |w\R & = E^{\pm \alpha}_n |w\R =0~, &\quad \quad \forall n>0~, 
\forall \alpha >0~,\\
H^i_n |w\R & = 0~,  & \quad \quad \forall n>0 ~,\forall i~,\\
H^i_0 |w\R & = (w)_i |w\R ~. & \\
\end{array}
\label{h.wt.state}  
\eea 
This state satisfies: $L^{SUG}_0 |w\R = h_{w} |w\R$ where $h_{w} = w^2/2 >0$. 
An irreducible representation labelled by $w$ is 
obtained by applying $J^{\mu \nu}_{-n}~~(n\geq0)$ on $|w\R$ in all possible 
ways. One checks unitarity i.e. the absence of the negative norm states with 
respect to the usual hermitian inner product. The representation obtained in 
this way contains null states which one has to 
put to zero by hand. For a unitary representation, the four dimensional vector 
defined by the $H^i_0$ $(i=1,...,4)$ eigenvalues of any state in the 
representation lies on a 
lattice $\Lambda_W$ called weight lattice which is dual to the root lattice
$\Lambda_R$ generated by all the nonzero roots. Since any root can be 
expanded in terms of the simple roots with integer coefficients, $\alpha^s_i$'s
 can be taken as the basis for $\Lambda_R$. It follows from the definition of a
lattice being dual to another, that $\Lambda_W$ can be generated by certain 
vectors $f_i$ called the fundamental weights defined by,
\beq
f_i.\alpha^s_j = \delta_{ij}~, \quad \quad 1\leq i~,j \leq 4 ~. 
\label{def.fund.wts.}
\eeq
Given eqns. $(\ref{roots})$ it is easy to obtain the following four $SO(8)$
fundamental weights:
\beq
f_1 =e\equiv e_1= \pmatrix{1\cr0\cr0\cr0}~, f_2 =\psi= 
\pmatrix{1\cr1\cr0\cr0}~,
f_3 =\bar{\delta} \equiv \frac{1}{2} \pmatrix{1\cr1\cr1\cr-1}~, 
f_4 =\delta \equiv \frac{1}{2} \pmatrix{1\cr1\cr1\cr1}~.
 \label{fund.wts.}
\eeq
Now it can be shown that the value of $k$ restricts the highest weights that 
can occur as labels of the representations. For $\hat{SO}(8)_{k=1}$, there are
four irreducible representations corresponding to the follwing highest 
weights:
\beq
w = 0~,\quad e~,\quad \bar{\delta}~,\quad \delta ~.   \label{h.wts.}
\eeq
It is easy to show that,
\beq
\Lambda_W = \Lambda_R \sqcup \Lambda_e \sqcup \Lambda_{\bar{\delta}}
\sqcup \Lambda_{\delta}   ~,\label{wt.latt.}
\eeq
where $\sqcup\equiv \hbox{union of disjoint sets}$ and $\Lambda_w \equiv w + 
\Lambda_R~~~~(w = e,~\bar{\delta},~\delta)$. 
 
Each highest weight representation corresponding to a weight $w$ has a finite dimensional subspace of states which are annihilated by $J^{\mu \nu}_n$ $\forall
n>0$. These form a representation of the zero mode subalgebra of the current algebra and will be called the vacuum representation corresponding to the given highest weight representation. If $n_{(w)}$ denotes the dimension of the vacuum
representation, we lable these states as $|\Pi^{(w)}_0, j\R$, $j=1,...,n_{(w)}$. These satisfy the following ptoperty:
\bea
J^{\mu \nu}_0 |{\Pi}^{(w)}_0,j \R &=& \lt( T_{(w)}^{\mu \nu} \rt)_{lj} 
|{\Pi}^{(w)}_0,l \R ~,     \label{def.vac.repns.}
\eea
where $T_{(w)}^{\mu \nu}$ are $n_{(w)} \times n_{(w)}$ hermitian 
matrices\footnote{For $SO(8)$ it is possible to impose the reality property on 
the representations. This makes the matrices $T_{(w)}^{\mu \nu}$ purely 
imaginary and antisymmetric.} representing the current zero-modes in the vacuum
representation and repeated indices are summed over. The $SO(8)$ group elements
 are parameterized in the following way:
\beq
{\cal O} \lt(\theta \rt) = \exp \lt( \frac{i}{2} \theta_{\mu \nu} 
J^{\mu \nu}_0 \rt)      ~,\label{so8rot.}
\eeq
where $\theta_{\mu \nu} = - \theta_{\nu \mu}$, 
${\theta^*_{\mu \nu}} = \theta_{\mu \nu}$. Clearly the vacuum states are 
created by applying the current zero-modes on the highest weight state. The 
explicit matrix representations for the four vacuum representations of 
$\hat{SO}(8)_{k=1}$ are given in appendix \ref{app-vac.rep}. 

\sectiono{Realization of $\hat{SO}(8)_{k=1}$ in terms of NSR, GS and Bosonic 
Fields}
\label{sec-realization}
The Euclidean actions in the NSR and GS formalism take the following form on 
the $z$-plane:
\bea
S_{NSR} &=& \frac{1}{4\pi} \int d^2z \lt(\Psi^{\mu} \bar{\partial}\Psi^{\mu} 
+  \tilde{\Psi}^{\mu} \partial \tilde{\Psi}^{\mu} \rt)   ~,\label{Snsr} \\
S^{IIB}_{GS} &=& \frac{1}{4\pi} \int d^2z \lt( S^a \bar{\partial} S^a 
+ \tilde{S}^a \partial \tilde{S}^a \rt)   ~,\label{S2bgs}\\
S^{IIA}_{GS} &=& \frac{1}{4\pi} \int d^2z \lt( S^a \bar{\partial} S^a 
+ \tilde{S}^{\dot{a}} \partial \tilde{S}^{\dot{a}} \rt)~.   \label{S2ags}
\eea
$\Psi^{\mu}(z)$ and $\tilde{\Psi}^{\mu}(\bar z)$ transform as the states
$|\mu\R$, $S^a(z)$, $\tilde{S}^a(\bar z)$ as $|a\R$ and $\tilde S^{\dot a}
(\bar z)$ as $|\dot a\R $ under $SO(8)$ with the transformation laws defined 
in appendix \ref{app-vac.rep}. All the holomorphic fields have conformal 
dimension $(\frac{1}{2},0)$ and antiholomorphic fields have conformal dimension 
$(0,\frac{1}{2})$. The mode expansions are:
\bea
\Psi^{\mu}(z) = \sum_{r\in {\bf Z}+\nu} \frac{\Psi^{\mu}_r}{z^{r+1/2}} ~,\quad
\tilde{\Psi}^{\mu}(\bar z) = \sum_{r\in {\bf Z}+\tilde{\nu}} 
\frac{\tilde{\Psi}^{\mu}_r}{\bar{z}^{r+1/2}}   ~,\label{psi.mode}
\eea
where $\nu, \tilde{\nu} = 0,\frac{1}{2}$ corresponding to respectively R and NS sector.
\beq
S^a(z) = \sum_{n\in {\bf Z}} \frac{S^a_n}{z^{n+1/2}}~,\quad
\tilde{S}^a(\bar z) = \sum_{n\in {\bf Z}} \frac{\tilde S^a_n}{\bar{z}^{n+1/2}}~,\quad   
\tilde{S}^{\dot a}(\bar z) = \sum_{n\in {\bf Z}} \frac{\tilde S^{\dot a}_n}
{\bar{z}^{n+1/2}} ~.    \label{S.mode}
\eeq  
The nontrivial anticommutation relations are:
\beq
\lt\{ \Psi^{\mu}_r, \Psi^{\nu}_s \rt\} = \delta^{\mu \nu} \delta_{r+s,0}~,
\quad
\lt\{ S^a_m, S^b_n \rt\} = \delta^{a b} \delta_{m+n,0}  ~,\label{anticommut}
\eeq
and similarly for the antiholomorphic modes. The OPE's are:
\beq
\Psi^{\mu}(z) \Psi^{\nu}(0) \sim \frac{\delta^{\mu \nu}}{z}~,~~~~
S^a(z)S^b(0) \sim \frac{\delta^{ab}}{z}    ~~,\label{psi-psi.S-S.ope}
\eeq
and similarly for the antiholomorphic fields. Now we define the fermionic part 
of the energy-momentum tensors in NSR and GS formalism as the following:
\bea
T^{NSR}(z) = -\frac{1}{2} :\Psi^{\mu} \partial \Psi^{\mu}:(z)~,&&
T^{GS}(z) = -\frac{1}{2} :S^a \partial S^a:(z)   ~,\label{Tnsrgs}
\eea
and similarly for the antiholomorphic part. For $\tilde{T}^{GS}(\bar z)$ we 
have sum 
over $a$ or $\dot a$ depending upon whether it is IIB or IIA respectively. The
 conformal normal ordering of the fermionic fields are defined as follows:
\bea
:\Psi^{\mu}\Psi^{\nu}(z): &=& \lim_{w\rightarrow z} \lt[ \Psi^{\mu}(w)
\Psi^{\nu}(z) - \frac{\delta^{\mu \nu}}{(w-z)} \rt] ~, \cr
:S^a S^b(z): &=& \lim_{w\rightarrow z} \lt[ S^a(w) S^b(z) - 
\frac{\delta^{ab}}{(w-z)} \rt]~.  \label{psi.S.norm.ord}
\eea
Each of the above energy-momentum tensors has standard OPE with itself with
central charge $c=4$, which is the same as that for $T^{SUG}$.

In addition to the NSR and GS formalisms we will consider another 
formalism, which is
 obtained through abelian bosonization\cite{bosonization} of the fermions. 
This is well known that
the theory of the four world-sheet bosonic fields obtained by bosonizing 
the eight world-sheet fermions (either of NSR and GS type) is 
equivalent to the fermionic theory, if the bosonic momenta are properly 
quantized. The reason we cosider the bosonic formalism is that the basis of
states in the Hilbert space that we will construct will have explicit 
resemblance with the natural basis 
that one would construct in the bosonic formalism. Here we will give the 
bosonization rules and the basic definitions in the 
bosonic formalism as we have done in the fermionic case. We start by defining 
the fermionic fields in the lattice basis:
\bea
\lambda^{\pm e_j}(z) &\equiv& \frac{1}{\sqrt 2} \lt( \Psi^{\mu =2j-1}(z) \mp 
i \Psi^{\mu =2j}(z) \rt) ~, \cr
\chi ^{\pm \delta_j}(z) &\equiv & \frac{1}{\sqrt 2} \lt( S^{a =2j-1}(z) \mp i
S^{a =2j}(z) \rt) ~, \cr
\tilde{\xi}^{\pm \bar{\delta_j}}(\bar{z}) &\equiv & \frac{1}{\sqrt 2} \lt( 
\tilde{S}^{\dot{a} =2j-1}(\bar{z}) \mp i \tilde{S}^{\dot{a} =2j}(\bar{z}) 
\rt)~.  
\label{latt.fields}
\eea
and the other antiholomorphic fields namely $\tilde{\lambda}^{\pm e_i}(\bar{z})
$ and $\tilde{\chi}^{\pm \delta_i}(\bar z)$ are defined in the similar fashion.
 Then we bosonize in the following way:
\bea
\lambda^{\pm e_i}(z) \sim :\exp \lt( \pm i e_i. \phi \rt) (z): ~,\cr
\chi^{\pm \delta_i}(z) \sim :\exp \lt( \pm i \delta_i. \phi \rt) (z):~,\cr
\tilde{\xi}^{\pm \bar{\delta_i}}(\bar{z}) \sim :\exp \lt( \pm i 
\bar{\delta}_i. \tilde{\phi} \rt) (\bar z):  ~.          
\label{bosonize}
\eea
Bosonizations of the other antiholomorphic fields are similar to the 
corresponding holomorphic fields with $\phi (z)$ replaced by $\tilde{\phi}
(\bar z)$. The symbol $\sim$ in the above equations implies that the relations 
are true upto proper cocycle factors which need to satisfy various consistency 
conditions. We will not need the explicit form of the cocycle factors for our 
analysis and will always mention whenever the symbol $\sim$ means it\footnote{
If not mentioned, it will always carry its usual meaning, for example in case 
of OPE, it would imply that the relation is true up to a regular part.}. The 
fields $\phi (z)$ and $\tilde \phi (\bar z)$ are four 
component fields representing the CSA space. These fields 
are compactified in such a way that the compact momenta take values on $\Lambda
_W$. The mode expansions for the bosonic fields and commutation relations 
for different modes can be shown to be as follows:
\bea
\phi_i(z)\equiv e_i.\phi(z) = \varphi^i -iP^i\ln z + i\sum_{n\neq 0} 
\frac{1}{n} \frac{\beta^i_n}{z^n} ~, \label{phi.mode} \\
\lt[\varphi^i,P^j\rt]=i \delta^{ij}~,\quad 
\lt[\beta^i_m,\beta^j_n\rt]= m \delta_{m+n,0}~ \delta^{ij}~. 
\label{commut}
\eea
The exponential operator in the bosonization rules has the standard definition 
in terms of the modes:  
\beq
:e^{i\rho .\phi}(z): \equiv e^{i\rho .\varphi} z^{\rho .P} e^{-\sum_{n<0} 
\frac{\rho.\beta_n}{nz^n}} e^{-\sum_{n>0} \frac{\rho .\beta_n}{nz^n}} ~.
\label{exp.phi}
\eeq
One can derive the OPE:
\beq
\phi^i(z) \phi^j(0) \sim - \delta^{ij} \ln z \label{phi-phi.OPE}~.
\eeq
The energy-momentum tensor is defined as,
\beq
T^{BOS}(z) = - \frac{1}{2} :\partial \phi^i \partial \phi^i:(z) \label{Tbos}~.
\eeq
where,
\bea
:\partial \phi^i \partial \phi^i:(z) = \lim_{w\rightarrow z} \lt[ \partial_w
\phi^i(w) \partial_z \phi^i(z) + \frac{4}{\lt(w-z\rt)^2} \rt] ~.
\label{bos.norm.ord}
\eea
Again $T^{BOS}(z)$ has the usual OPE with itself with central charge $c=4$. For
 the right moving part there is a similar eqn. for each of the above eqns. 
(\ref{phi.mode}) - (\ref{bos.norm.ord}). Now let us specify the $SO$(8) 
currents in different formalisms:\\

{\bf NSR:}
\bea
J^{\mu \nu}(z) = \frac{i}{2} \Sigma^{\mu \nu}_{\rho \sigma} 
:\Psi^{\rho}\Psi^{\sigma}(z): &,& \tilde J^{\mu \nu}(\bar z) = \frac{i}{2} 
\Sigma^{\mu \nu}_{\rho \sigma} :\tilde{\Psi}^{\rho}
\tilde{\Psi}^{\sigma}(\bar z):~,   \label{Jnsr}
\eea
where\footnote{Upper and lower indices are indistinguishable for all the 
fundamental representations.} 
$\Sigma^{\mu \nu}_{\rho \sigma} = \delta^{\mu \rho} \delta^{\nu \sigma} -
 \delta^{\mu \sigma} \delta^{\nu \rho}$.\\

{\bf GS:}
\bea
\begin{array}{lll}
J^{\mu \nu}(z) = &
\frac{i}{4} \lt(\gamma^{\mu}\bar{\gamma}^{\nu}\rt)_{ab} :S^a S^b(z): &  
\quad \lt(\hbox{IIB, IIA} \rt)~, \\
&&\\
\tilde{J}^{\mu \nu}(\bar{z}) =& 
\frac{i}{4} \lt(\gamma^{\mu} \bar{\gamma}^{\nu}\rt)_{ab} 
:\tilde{S}^a \tilde{S}^b(\bar{z}): &
\quad \lt( \hbox{IIB} \rt) ~,\\
&&\\
\tilde{J}^{\mu \nu}(\bar{z}) =&
\frac{i}{4} \lt(\bar{\gamma}^{\mu}{\gamma}^{\nu}\rt)_{\dot a \dot b} 
:\tilde{S}^{\dot a} \tilde{S}^{\dot b}(\bar{z}): &
\quad \lt( \hbox{IIA} \rt)~.
\end{array}   
\label{Jgs}
\eea
where $\gamma^{\mu}$, $\bar{\gamma}^{\mu}$ are $8 \times 8$ matrices 
defined in appendix \ref{app-gamma}.\\

{\bf Bosonic:}
\bea
\begin{array}{ll}
H^i(z) = i \partial \phi^i(z)~, & \quad
E^{\alpha}(z) \sim~:e^{i\alpha .\phi}(z):~,\\
\tilde H^i(\bar z) = i \bar{\partial} \tilde{\phi}^i(\bar z)~,& \quad
\tilde E^{\alpha}(\bar z) \sim~:e^{i\alpha .\tilde \phi}(\bar z):~.
\end{array}
\label{HEbos}
\eea
Here the symbol $\sim $ indicates the presence of a suitable cocycle factor.
Note that $\beta^i_n=H^i_n$ and $P^i=H^i_0$. Now given all these, one can 
explicitly check that in a given formalism (NSR/
GS/ Bosonic) the fundamental fields have the same OPE with $T^{SUG}$ and the 
$T$ defined in that formalism. Also $T^{SUG}$ and all the other energy-momentum
tensors have the same central charge $c=4$. This establishes the equivalence of
all the energy-momentum tensors. In fact we can write:
\beq
T^{NSR} = T^{GS} = T^{BOS} \simeq T^{SUG} ~,     \label{Tequiv}
\eeq
where the symbol $\simeq$ implies that the action of $T^{SUG}$ on any state 
differs from that of $T^{NSR}$, $T^{GS}$ or $T^{BOS}$ only by a null state 
written in terms of the current modes. Here it should be noted that the null 
states become identically zero when expressed in terms of the fermionic or the 
bosonic variables, because these realizations are free of null states.

Now we will discuss how type IIA and IIB theories differ as realizations of 
the same current algebra $\hat{SO}(8)_{k=1}$. To see the difference it is 
easiest to consider the GS formalism. Let us consider IIB theory in this 
formalism. Recall that the 
field content is $S^a(z)$, $\tilde{S}^a(\bar{z})$ $(a=1,2,..,8)$ with integer 
modes for all fields. Because of the existence of zero modes $S^a_0$, 
$\tilde{S}^a_0$ and their specific anticommutation relations 
(eqn.~(\ref{anticommut})), the vacuum states
 (which get annihilated by all the +ve mode oscillators) have the structure 
\cite{gsw}
$$ \pmatrix{|\dot a\R \cr |\mu \R} \otimes 
\pmatrix{\widetilde{|\dot a\R} \cr \widetilde{|\mu \R}}~. $$
(Transformations of these states under $SO(8)$ are given in appendix 
\ref{app-vac.rep}.)
Since both the left and right sectors have the same structure, we analyze the 
spectrum built up only on one sector. The spectrum is obtained by acting 
$\chi_n^{\pm \delta_i}~~(n<0,i=1,..,4)$ in all possible ways on the vacuum 
states, where the modes $\chi_n^{\pm \delta_i}$ are defined by taking the same 
linear combinations as in eqns.(\ref{latt.fields}). One can show that the 
$SO(8)$ weight vectors corresponding to the states obtained in this way can be 
written in the following way:
\bea
P = \sum_i n^{\delta}_i \delta_i + e_1~, \quad \quad \quad n^{\delta}_i \in 
{\bf Z}~, 
\label{P.delta}
\eea
where $e_1$ and $\delta_i$ are defined in eqns.~(\ref{e}) and (\ref{d}) 
respectively.
It should be noted that $e_1$ here, is merely a choice. $P$ takes the same form
(\ref{P.delta})
if we replace $e_1$ by any weight in $\Pi_0^{(e)}$ or $\Pi_0^{(\bar{\delta})}$
 (see appendix \ref{app-vac.rep}). Now using the explicit form of the roots 
(eqn.(\ref{roots})), the weights
$\delta_i$ (eqn.(\ref{d})) and $\bar{\delta}_2$ (eqn.(\ref{db})), one can 
show that 
\begin{enumerate}
\item
When $\displaystyle{\sum_i n^{\delta}_i} =$ odd,
\bea
P = P^{odd} = \sum_i \bar{n}^s_i \alpha^s_i + \bar{\delta}_2 ~,
\label{P.alpha.s1}
\eea
where $\bar{n}^s_i \in {\bf Z}$. Furthermore given any set of integer values 
for 
$\{\bar{n}^s_i\}$ in (\ref{P.alpha.s1}), we can find integers 
$\lt\{n^{\delta}_i \rt\}$ satisfying
eqns.~(\ref{P.delta}) and (\ref{P.alpha.s1}) with 
$\sum_i n^{\delta}_i =$ odd.
\item
When $\displaystyle{\sum_i n^{\delta}_i} =$ even,
\bea
P = P^{even} = \sum_i n^s_i \alpha^s_i + e_1 ~,
\label{P.alpha.s2}
\eea
where $n^s_i \in {\bf Z}$. Also given any set of integer values for 
$\{ n^s_i \}$ in (\ref{P.alpha.s2}), $\{n^{\delta}_i \}$ defined through 
eqns.~(\ref{P.delta}) and (\ref{P.alpha.s2}), takes integer values with the 
condition $\sum_i n^{\delta}_i =$ even.
\end{enumerate}
These imply that the set of $SO(8)$ weight vectors $\{P\}$ in the spectrum is 
the union of two disjoint subsets namely $\{P^{even}\}$ and $\{P^{odd}\}$, and 
they cover fully the two parts of the weight lattice: $\Lambda_e$ and 
$\Lambda_{\bar{\delta}}$ respectively. Therefore the spectrum is the direct sum
of $\Pi^{(e)}$ and $\Pi^{(\bar{\delta})}$, where $\Pi^{(w)}$ is the full 
irreducible representation built up over the vacuum representation 
$\Pi^{(w)}_0$. Hence the full spectrum\footnote{We remind the reader that here
we are dealing with the world-sheet fermionic part only.} of the IIB theory 
can be expressed in terms of the irreducible components of $\hat{SO}(8)_{k=1}$ 
in the following way:
\beq
\hbox{type~IIB}\quad : \quad \quad \lt(\Pi^{(e)} \oplus 
\Pi^{(\bar{\delta})}\rt) \otimes \lt(\Pi^{(e)} \oplus \Pi^{(\bar{\delta})}\rt) 
~.
\label{IIB}
\eeq
In type IIA theory the field content for the left sector is the same as that 
in type IIB, but for the right sector $\tilde S^a(\bar z)$ is replaced by  
$\tilde S^{\dot a}(\bar z)$. Similar analysis shows that the spectrum of IIA 
theory, in terms of irreducible components of $\hat{SO}(8)_{k=1}$, takes the 
following form:
\beq
\hbox{type~IIA}\quad : \quad \quad \lt(\Pi^{(e)} \oplus 
\Pi^{(\bar{\delta})}\rt) \otimes \lt(\Pi^{(e)} \oplus \Pi^{(\delta)}\rt) ~.  
\label{IIA}
\eeq   
At this point one notices that $\Pi^{(0)}$ is absent in both the theories. 
Similar analysis can be done in the NSR formalism. Thus for example in IIB, the
NS sector states after GSO projection belong to $\Pi^{(e)}$ and R sector 
states, after GSO projection, belong to $\Pi^{(\bar \delta)}$.

Now a final comment about the fermionic realizations: the current algebraic 
vacuum states $|\mu\R,~|a\R,~|\dot a\R$ appear as representations of various 
zero-mode oscillators of IIA and IIB theories in different realizations. Thus 
the latter states are same as the current algebraic vacuum states up to some 
overall phases. We will take this phase to be $1$ in each case.

\sectiono{Ishibashi States and Type IIB D-Branes}
\label{sec-Ishib}
\subsectiono{Ishibashi States}
\label{subsec-bais} 
Let us consider a two dimensional conformal field theory on the complex plane,
 both the left and right parts of which realize the same current algebra $\hat
g$. If $J^a(z)$ and $\tilde J^a(\bar z)$ are the local currents (in the 
covariant basis) with modes $J^a_n$ and $\tilde J^a_n$ respectively, then one 
considers certain boundary conditions connecting $J^a(z)$ and $\tilde J^a
(\bar z)$ at the boundary ($|z|=|\bar z|=1$)\footnote{On the cylinder, $t=0$
 is cosidered to be the boundary which under the conformal transformation 
$z= e^t e^{-i\sigma}$, gets mapped on to the unit circle $|z|=|\bar z|=1$
on the $z$-plane.} which can be written in terms of 
their modes in the following way:
\beq
J^a_n + \tau \lt(\tilde J^a_{-n} \rt) = 0 ~. \label{J.b.c}
\eeq  
Here $\tau$ is an inner or outer automorphism of $\hat g$. An Ishibashi state 
is a solution of the above boundary condition and is given by \cite{ko}:
\beq
|w \R\R = \sum_N |N,w\R \otimes {\cal T} \Theta \widetilde{|N,w{\R}} ~, 
\label{def.Ishb}
\eeq 
where $\lt \{ |N,w\R \rt \}$ is a complete set of orthonormal basis vectors for
 an unitary irreducible representation of $\hat g$ labelled by the highest
weight $w$. $\Theta$ is an antiunitary operator on the Hilbert space defined by
the following action:
\bea
\Theta \tilde J^a_n \Theta^{-1} &=& - \tilde J^a_n ~,\cr
\Theta \widetilde{|w\R} &=& \widetilde{|-w\R} ~,   \label{theta.act}
\eea
where $\widetilde{|w\R}$ is the highest weight state defined in eqns. 
(\ref{h.wt.state}). 
${\cal T}$ is the Hilbert space operator corresponding to the automorphism 
$\tau$, and acts only on the right part:
\beq
{\cal T} \tilde J^a_n {\cal T}^{-1} = \tau \lt( \tilde J^a_n \rt)~. 
\label{calT.act}
\eeq
One has to properly take into account the action of ${\cal T}$ on $|w\R$ also. 
It has been shown in ref.\cite{ko} that the fact that $|w \R\R$ satisfies 
the boundary condition (\ref{J.b.c}) is independent of the choice of the 
orthonormal basis $\lt \{ |N,w\R \rt \}$.
\subsectiono{Type IIB D-Branes}
\label{subsec-IIBbranes}
We will use the Ishibashi states defined above to derive all the D-brane 
boundary states in type IIB. There are two points involved - the boundary 
conditions given by $\tau$ (eqn.~(\ref{J.b.c})) or equivalently the Hilbert 
space operator ${\cal T}$ (eqn.~(\ref{calT.act})) and the explicit construction
 of the basis. From eqn.~(\ref{def.Ishb}) it is clear that after constructing 
the basis one has to implement the action of ${\cal T}\Theta$ on the basis 
states (only right part). As has been discussed in sec.~\ref{sec-int} we will 
consider ${\cal T}$ to be corresponding to the reflection along a set of 
directions.

Now we will say something about the basis construction. The boundary state 
does not depend on the specific basis chosen.
The basis can be given purely in terms of the current modes or since the 
current algebra has different realizations, it could be given in terms of the 
variables
 of a specific realization. A basis which looks simple in one realization may
have very complicated expression in some other realization. We will show that 
choosing the basis properly in NSR and GS formalisms we can make the 
expressions for the BPS boundary states look exactly like the already existing 
solutions for those in the two formalisms. We will also show that the chosen
 basis in NSR is good enough for expressing also the non-BPS boundary states in
 that formalism. This however is not true in case of the GS formalism. We will 
explain why one needs to consider a different basis for expressing the non-BPS 
branes in GS formalism.\\

\noindent{\bf Current Algebraic derivation of BPS results of IIB in NSR and GS 
formalisms:}\\
Let us consider a BPS brane $Dp$ for $p$ = odd. We specify its 
alignment by an ordered set of indices $\lt\{ \mu_1, \mu_2,\cdots ,\mu_{p+1} 
\rt\}$ with the following chosen ordering: $ \mu_1< \mu_2<\cdots <\mu_{p+1}$.
Before we go into the derivation of the corresponding 
boundary state, we give some definitions and fix some conventions that are 
needed for the derivation. We define the following matrices \cite{gg} 
corresponding to the above mentioned ordered set of indices:
\bea
M_{\mu \nu} &=& \lt \{ \begin{array}{ll} \delta_{\mu \nu}\quad & \mu \notin 
\lt\{ \mu_i |i= 1,\cdots ,p+1 \rt\}~,\quad \nu = 1,2,\cdots ,8~, \\
-\delta_{\mu \nu} \quad & \mu \in \lt\{ \mu_i |i=1,\cdots ,p+1 \rt\}~,\quad 
\nu =1,2,\cdots ,8~, \end{array}\rt. \cr
M_{ab} &=& \lt( \gamma^{\mu_1} \bar{\gamma}^{\mu_2}\gamma^{\mu_3} 
\bar{\gamma}^{\mu_4}\cdots \gamma^{\mu_p}\bar{\gamma}^{\mu_{p+1}}\rt)_{ab}~, 
\cr
M_{\dot a\dot b} &=& \lt( \bar{\gamma}^{\mu_1} \gamma^{\mu_2}
\bar{\gamma}^{\mu_3} \gamma^{\mu_4}\cdots \bar{\gamma}^{\mu_p}
\gamma^{\mu_{p+1}} \rt)_{\dot a\dot b}~. 
\label{Mmatrices} 
\eea
Following are some useful properties satisfied by the above matrices \footnote{
 We could change the definition of $M_{ab}$ and $M_{\dot a\dot b}$ by a ($-$) 
sign without affecting these properties, but this ambiguity does not affect the
 physical results.}:
\bea
&M_{ab} M_{cb} = \delta_{ac}~,\quad
M_{\dot a\dot b} M_{\dot c\dot b} = \delta_{\dot a\dot c} ~,& \cr
&\gamma^{\mu}_{a\dot a} - M_{\dot a\dot b} M_{ab} M_{\mu \nu} \gamma^{\nu}_
{b\dot b} = 0 ~.& \label{Mprop}
\eea 
These matrices are respectively the $SO(8)$ vector, spinor and conjugate spinor
 representations of the reflections along $\mu_1, 
\mu_2, ...,$ $\mu_{p+1}$. Below we enlist the conventions that we will follow:
\begin{enumerate}
\item The 16 dimensional representaion, (spinor $\oplus$ conjugate spinor) 
for $\Psi^{\mu}_0$, is obtained by applying  various $\Psi^{\mu}_0$ oscillators
 on the highest weight state 
$|\delta \R$ which is a spin field acting on the SL(2,{\bf C}) vacuum. Our 
convention
 is to take this spin field to be anticommuting with the right moving fields 
$\tilde \Psi^{\mu}$. Now even number of $\Psi^{\mu}_0$ oscilltors act on 
$|\delta \R$ to give spinor representation and odd number to give conjugate 
spinor 
representation. Therefore $\tilde \Psi^{\mu}_0$ goes through $|\dot a\R$ 
 and picks up a ($-$) sign for $|a\R$. Due to the similar convention in GS 
formalism, $\tilde S^a_0$ passes through $|\dot a\R$ and picks up a ($-$) sign 
for $|\mu \R$. 
\label{conv.1}
\item The GSO rules for the R sector vacuum states are taken to be:
\bea
\lt(-1\rt)^f \pmatrix{|a\R \cr |\dot a\R}&=&\pmatrix{-|a\R \cr|\dot a\R}~,
\label{GSOrules}
\eea
where $f$ is the left moving world-sheet fermion number. Since we are 
considering IIB, the GSO rules for the right part is also the same.
\label{conv.2}
\end{enumerate}
Let us first derive the NSR BPS states 
through explicit basis construction. The R-R part of the boundary state 
corresponds to the irreducible representation $\lt( \Pi^{(\bar \delta)} 
\otimes \Pi^{(\bar \delta)} \rt) $ (eqn.(\ref{IIB})). We choose the following 
basis (eqn.~(\ref{def.Ishb})) in $\Pi^{(\bar \delta)}$ for the left part:
\bea
\lt \{|N,\bar \delta \R \rt \} & \equiv &
\lt\{ \lt[\prod_{\stackrel{n>0,}{\mu}} \lt(\Psi^{\mu}_{-n} \rt)^{N_{\mu n}} 
|\dot a\R \rt]_{\sum N_{\mu n} = even}~, \rt.\cr
&& \lt. \quad
\lt[\prod_{\stackrel{n>0,}{\mu}} \lt(\Psi^{\mu}_{-n}  \rt)^{N_{\mu n}} |a\R 
\rt]_{\sum N_{\mu n} =odd} \rt\} ~,
\label{N.deltabar.basis}
\eea
where $N_{\mu n}=0,1$ and the products are ordered products with some chosen 
ordering. The states $|a\R$ are absent in the representation 
$\Pi^{(\bar \delta)}$ and should be 
understood as being created by applying the $\Psi$ zero-modes on the states 
$|\dot a\R$. We may define these in the following way:
\bea
|a\R = \sqrt 2 \gamma^{\mu}_{a\dot b} \Psi^{\mu}_0 |\dot b \R ~.
\label{def.a}
\eea
The basis of states on the right part is defined similarly by replacing the
left handed oscillators by the right handed ones in eqn. 
(\ref{N.deltabar.basis}). The states which appear in the Ishibashi state are:
\bea
\lt \{|N,\bar \delta \R \otimes \widetilde{|N,\bar \delta \R} \rt \} 
& = & 
\lt \{ \left|\{N_{\nu n}\}_{(even)}, \dot a\right\R ~, \quad 
\left|\{N_{\nu n}\}_{(odd)}~, a\right\R \rt \} ~,   \label{Rnsr.basis}
\eea 
where, 
\bea
\left|\{N_{\mu n}\}_{(even)}, \dot a\right\R &\equiv& 
\lt[\prod_{\stackrel{n>0,}{\mu}} \lt(\Psi^{\mu}_{-n}  \rt)^{N_{\mu n}} 
|\dot a\R \otimes
\prod_{\stackrel{n>0,}{\mu}}\lt(\tilde{\Psi}^{\mu}_{-n}\rt)^{N_{\mu n}} 
\widetilde{|\dot a\R} \rt]_ {\sum N_{\mu n} =even}~,  \cr
&=& \lt[ \prod_{\stackrel{n>0,}{\mu}} \lt(i \Psi^{\mu}_{-n} \tilde 
\Psi^{\mu}_{-n}
\rt)^{N_{\mu n}} |\dot a\R \otimes \widetilde{|\dot a \R}  
\rt]_{\sum N_{\mu n} =even} ~.
\label{def.Rnsr.basis1}
\eea
\bea
\left|\{N_{\mu n}\}_{(odd)}, a\right\R &\equiv& 
\lt[\prod_{\stackrel{n>0,}{\mu}} \lt(\Psi^{\mu}_{-n}  \rt)^{N_{\mu n}} 
|a\R  \otimes
\prod_{\stackrel{n>0,}{\mu}}\lt(\tilde{\Psi}^{\mu}_{-n}\rt)^
{N_{\mu n}} \widetilde{|a\R} \rt]_ {\sum N_{\mu n} =odd} ~,  \cr
&=& i \lt[ \prod_{\stackrel{n>0,}{\mu}} \lt(i \Psi^{\mu}_{-n} \tilde 
\Psi^{\mu}_{-n}
\rt)^{N_{\mu n}} |a\R \otimes \widetilde{|a\R}  
\rt]_{\sum N_{\mu n} =odd} ~.
\label{def.Rnsr.basis2} 
\eea
Here we have used the anticommutation rule of $\Psi^{\mu}_n$ with 
$\tilde \Psi^{\nu}_m$, and the fact that $\tilde \Psi^{\mu}_n$ passes through 
$|\dot a\R$ without picking any $(-)$ sign, and picks up a $(-)$ sign while 
passing
 through $|a\R$ (see convention \ref{conv.1}). Now we have to find the action 
of $\Theta$ on the states in eqns.~(\ref{def.Rnsr.basis1}) and 
(\ref{def.Rnsr.basis2}). Using eqns.~(\ref{theta.act}), (\ref{Jnsr}), 
(\ref{pmdbi}), (\ref{pdb2.rep.}) and (\ref{mdb2.rep.}) and the fact that 
$\Theta$ is antilinear, one derives:
\bea
\Theta \widetilde{|\dot a\R} & = & \widetilde{|\dot a\R}~, \nonumber\\
\Theta \tilde{\Psi}^{\mu}_n \Theta^{-1} & = & \tilde{\Psi}^{\mu}_n~.
\label{theta.act.Psi}
\eea
This with eqns.~(\ref{def.a}), (\ref{def.Rnsr.basis1}) and  
(\ref{def.Rnsr.basis2}) says that all the states in (\ref{Rnsr.basis}) are 
invariant under $\Theta$. For the $Dp$-brane that we are considering, we have 
the following action of ${\cal T}$ (eqn \ref{def.Ishb}):
\bea
\begin{array}{c}
{\cal T} \tilde{\Psi}^{\mu}_n {\cal T}^{-1} = M_{\mu \nu} 
\tilde{\Psi}^{\nu}_n ~,\\
{\cal T}\widetilde{|a\R} = M_{a b}\widetilde{|b\R}~, \quad
{\cal T}\widetilde{|\dot a\R} = M_{\dot a \dot b}\widetilde{|\dot b\R}~.  
\end{array}
\label{calT.act.Psi}
\eea
Using eqns.~(\ref{def.Rnsr.basis1}), (\ref{def.Rnsr.basis2}) and 
(\ref{calT.act.Psi}) and the fact that $\Theta$ acts trivially on the states 
in (\ref{Rnsr.basis}) we get:
$$
\begin{array}{rr}
{\cal T} \Theta \displaystyle{\sum_{\{ N_{\mu n}\},\dot a }^{even}} 
\left|\{N_{\mu n}\}_{(even)},\dot a\right\R & = 
\displaystyle{\sum_{\{ N_{\mu n}\},\dot a,\dot b}^{even}}~ 
\prod_{\stackrel{n>0,}{\mu}} 
\lt(i\Psi^{\mu}_{-n} \lt(\sum_{\nu} M_{\mu \nu} \tilde{\Psi}^{\nu}_{-n}\rt) 
\rt)^{N_{\mu n}} \\
& M_{\dot a \dot b} |\dot a\R \otimes \widetilde{|\dot b \R}~, \\
& \\
{\cal T} \Theta \displaystyle{\sum_{\{ N_{\mu n}\}, a}^{odd}}
 \left|\{N_{\mu n}\}_{(odd)}, a\right\R & = 
i\displaystyle{\sum_{\{ N_{\mu n}\},a,b}^{odd}}~ 
\prod_{\stackrel{n>0,}{\mu}} 
\lt(i\Psi^{\mu}_{-n} \lt(\sum_{\nu} M_{\mu \nu} \tilde{\Psi}^{\nu}_{-n}\rt) 
\rt)^{N_{\mu n}} \\
&  M_{ab} |a\R \otimes \widetilde{|b \R}~, 
\end{array}
$$
\bea
\label{calT.Theta.act}
\eea
where $\displaystyle{\sum_{\{ N_{\mu n}\}}^{even}}$ and 
$\displaystyle{\sum_{\{ N_{\mu n}\}}^{odd}}$ imply sum over the set 
$\lt\{N_{\mu n}\rt\}$ of numbers $N_{\mu n}=0,1$ with 
$\displaystyle{\sum_{\stackrel{n>0,}{\mu}}}N_{\mu n}$ = even and odd 
respectively. 
Then using eqns.~(\ref{def.Ishb}), (\ref{Rnsr.basis}) and  
(\ref{calT.Theta.act}) it is straightforward to show that the Ishibashi state
$|p, \bar \delta~\R\R$ corresponding to the specific $D$-brane we are 
considering is given by:
\bea
|p, \bar \delta~\R\R = |p,RR \R_{NSR}~,   
\eea
where 
\bea
|p,RR\R_{NSR} &=& \frac{1}{2} \lt[ |p,RR,+\R + |p,RR,-\R \rt]~, \nonumber\\
|p,RR,\pm \R &=& \exp\lt[ \pm i \sum_{n>0} \Psi^{\mu}_{-n}
M_{\mu \nu} \tilde{\Psi}^{\nu}_{-n}\rt] |p,RR,\pm \R_0 ~, \nonumber\\
|p,RR,\pm \R_0 &=& M_{\dot a \dot b} |\dot a\R \otimes \widetilde{|\dot b \R}
\pm i M_{ab} |a\R \otimes \widetilde{|b \R} ~.     \label{RRnsr}    
\eea
According to the convention adopted for the GSO rules, $|p,RR\R_{NSR}$
is GSO invariant. One can show that the R-R part of the actual $Dp$ brane 
boundary state written in the NSR formalism takes the form 
${\cal A} |p,RR \R_{NSR}$, where the operator ${\cal A}$ does not involve any
fermionic oscillator.

For the NS-NS part of the boundary state one starts with the irreducible 
representation $\lt( \Pi^{(e)} \otimes \Pi^{(e)} \rt)$ (eqn.(\ref{IIB}))
and proceeds in the same way. For constructing $|p,e~\R\R$ we choose the basis 
states to be,
\bea
\lt\{|N,e\R \rt\} \equiv \lt\{ \lt[\prod_{\stackrel{r>0,}{\mu}} \lt(
\Psi^{\mu}_{-r}  \rt)^{N_{\mu r}} |0\R \rt]_ {\sum N_{\mu r}=odd} \rt\} ~.
\label{NSnsr.basis}
\eea
The final result is 
\bea
|p, e~\R\R =  |p,NSNS \R_{NSR}~,
\eea
where 
\bea
|p,NSNS\R_{NSR} &=& \frac{1}{2} \lt[ |p,NSNS,+\R - |p,NSNS,-\R \rt]~,
\nonumber\\
|p,NSNS,\pm \R &=& \exp\lt[ \pm i \sum_{r>0} \Psi^{\mu}_{-r}
M_{\mu \nu} \tilde{\Psi}^{\nu}_{-r}\rt] |0\R \otimes \widetilde{|0\R}~. 
\label{NSNSnsr}
\eea
Again one can show that the  NS-NS part of the actual $Dp$ brane boundary 
state written in the NSR formalism takes the form ${\cal A} |p,NSNS\R_{NSR}$,
with the same operator ${\cal A}$ appearing in the R-R part. Therefore
$|p,\eta \R \equiv |p, e~\R\R + i\eta~|p, \bar \delta~\R\R$ reproduces 
the known expression for the $Dp$ ($\bar Dp$)-brane boundary state for 
$\eta=1$ ($\eta=-1$) in the sense that the complete $Dp$-brane boundary state 
takes the form ${\cal A} |p,\eta \R$.

For deriving the BPS boundary state in GS formalism we take the following 
basis:
\bea
\lt\{|N,e \R \rt\} &\equiv &
\lt\{ \lt[\prod_{a,n>0} \lt(S^a_{-n} \rt)^{N_{an}} |\mu \R 
\rt]_{\sum N_{a n} =even}~, \rt.\nonumber\\
&& \lt. \quad 
\lt[\prod_{a,n>0} \lt( S^a_{-n}  \rt)^{N_{an}} |\dot a\R 
\rt]_{\sum N_{an} =odd} \rt\} ~,
\label{NSgs.basis}
\eea
\bea
\lt\{|N,\bar \delta \R \rt\} &\equiv &
\lt\{ \lt[\prod_{a,n>0} \lt( S^a_{-n} \rt)^{N_{an}} |\dot a \R
 \rt]_{\sum N_{an} =even}~, \rt.\nonumber\\
&& \lt. \quad 
\lt[\prod_{a,n>0} \lt(S^a_{-n}  \rt)^{N_{an}} |\mu \R 
\rt]_{\sum N_{an} =odd} \rt\} ~,
\label{Rgs.basis}
\eea
and similarly for the right part. Then proceeding in the same way as described 
above one can show\footnote{One can again show that $\Theta$ acts trivially on
all the right moving basis states. The matrices $M_{\mu \nu}$,
$M_{ab}$ and $M_{\dot a \dot b}$ get introduced in the same way, as we saw in 
the NSR case, through the action of ${\cal T}$.},
\bea 
|p,\eta \R &=& \exp\lt[ -i \eta \sum_{n>0} S^a_{-n} M_{ab}
\tilde S^b_{-n} \rt] |p,\eta\R_0 ~, \nonumber\\
|p,\eta\R_0 &=& M_{\mu \nu} |\mu\R \otimes \widetilde{|\nu \R} + i \eta 
M_{\dot a\dot b} |\dot a\R \otimes \widetilde{|\dot b \R} ~. \label{BPSgs}
\eea
The above form of $|p,\eta \R$ agrees with the known expression \cite{gg} in GS
formalism\footnote{In the sense that the complete boundary state is 
${\cal A}|p,\eta \R$.}.\\ 

\noindent {\bf Non-BPS boundary states of IIB in NSR formalism and problem 
with GS formalism:}\\
Let us consider a non-BPS brane $Dp$ ($p$ even) in IIB with the alignment
given by the ordered set of indices $\lt\{\mu_1, \mu_2,\cdots ,\mu_{p+1}
\rt\}_{\mu_1<\mu_2<\cdots ,<\mu_{p+1}}$. The vector, spinor and conjugate 
spinor representations of the parity transformation along all of these 
directions are given respectively by:
\bea
\bar M_{\mu \nu} &=& \lt \{ \begin{array}{ll} \delta_{\mu \nu}\quad & \mu 
\notin \lt\{ \mu_i |i= 1,\cdots,p+1 \rt\} ~, \nu =1,2,\cdots,8~, \\
-\delta_{\mu \nu} \quad & \mu \in \lt\{ \mu_i |i=1,\cdots,p+1 \rt\} ~,
\nu =1,2,\cdots,8~,\end{array}
\rt. \nonumber\\
\bar M_{a\dot b} &=& \lt( \gamma^{\mu_1} \bar{\gamma}^{\mu_2}
\gamma^{\mu_3} \bar{\gamma}^{\mu_4}\cdots \gamma^{\mu_{p-1}} 
\bar{\gamma}^{\mu_p}\gamma^{\mu_{p+1}} \rt)_{a\dot b}~, \nonumber\\
\bar M_{\dot ab} &=& \lt( \bar{\gamma}^{\mu_1} \gamma^{\mu_2} 
\bar{\gamma}^{\mu_3} \gamma^{\mu_4} \cdots \bar \gamma^{\mu_{p-1}} 
\gamma^{\mu_p}\bar \gamma^{\mu_{p+1}} \rt)_{\dot ab} ~. 
\label{barMmatrices} 
\eea
The properties corresponding to that of eqns.~(\ref{Mprop}) satisfied by the 
$\bar M$ matrices\footnote{As before, the matrices 
$\bar M_{a\dot b}$ and $\bar M_{\dot ab}$ could be defined with a factor of 
($-$) sign without affecting these properties. But again this ambiguity is 
not physically relevant.} are,
\bea
&\bar M_{a\dot b} \bar M_{c\dot b} = \delta_{ac}~,\quad
\bar M_{\dot ab} \bar M_{\dot cb} = \delta_{\dot a\dot c} ~,& \nonumber\\
&\bar \gamma^{\mu}_{\dot aa} + \bar M_{\dot ab} \bar M_{a\dot b} 
\bar M_{\mu \nu} \bar \gamma^{\nu}_{\dot bb} = 0 ~.& 
\label{Mbarprop}
\eea 
We will discuss the current algebraic derivation of the corresponding boundary 
state in NSR formalism first. Let us first notice that the Ishibashi state
$|p,\bar \delta~\R\R$ corresponding to the alignment described above is not
realized in IIB. This is because the states belonging to the conjugate spinor 
representation labelled by the highest weight $\bar \delta$ go over to the 
states in the spinor representation labelled by the highest weight 
$\delta$ under the action of ${\cal T}$ correponding to this non-BPS boundary
condition\footnote{A reflection along odd number of directions corresponds to 
an outer automorphism of the $\hat{SO}(8)_k$ current algebra which interchanges
 the spinor and conjugate spinor representations but maps the vector 
representation to itself.\label{nbps.realize}}. Therefore we have:
\bea 
{\cal T}:\quad \Pi^{(\bar \delta)} \otimes \Pi^{(\bar \delta)}
\rightarrow \Pi^{(\bar \delta)} \otimes \Pi^{(\delta)}~.
\eea
But the states in $\lt(\Pi^{(\bar \delta)} \otimes \Pi^{(\delta)}\rt)$ are not 
present in IIB. The Ishibashi state $|p,e~\R\R$ does get
realized in IIB (see footnote \ref{nbps.realize}) and can be obtained by 
starting with the same basis (eqn. \ref{NSnsr.basis}) used in the BPS case.
Using the following action of ${\cal T}$:
\bea
{\cal T} \widetilde{|0\R} &=& \widetilde{|0\R}~,   \nonumber\\
{\cal T} \tilde \Psi^{\mu}_n {\cal T}^{-1} &=& \bar M_{\mu \nu} \tilde \Psi^
{\nu}_n~.
\label{calT.act.Psi2}
\eea
one gets,
\bea
|p,e~\R\R &=& \frac{1}{2} \lt[ |p,NSNS,+\R - |p,NSNS,-\R \rt]~,  \nonumber\\
|p,NSNS,\pm \R &=& \exp\lt[ \pm i \sum_{r>0} \Psi^{\mu}_{-r}
\bar M_{\mu \nu} \tilde{\Psi}^{\nu}_{-r}\rt] |0\R \otimes \widetilde{|0\R} ~.
\eea
As in the BPS case, the state $|p,e~\R\R$ agrees with the known 
expression for the non-BPS boundary state in the sense that the actual 
non-BPS boundary state has the form $\bar {\cal A}|p,e~\R\R$, where 
$\bar {\cal A}$ is an operator which does not contain the fermionic 
oscillators.

This procedure does not go through so smoothly in the GS formalism. Here one 
is supposed to start with the basis (\ref{NSgs.basis}) and apply 
${\cal T}\Theta$ on the right part. One would naively think that ${\cal T}$ has 
the following action:
\bea
{\cal T} \widetilde{|\mu \R} &=& \bar M_{\mu \nu} \widetilde{|\nu \R}~, 
\nonumber\\
{\cal T} \widetilde{|\dot a \R} &=& \bar M_{\dot ab} \widetilde{|b \R} ~,
\nonumber\\
{\cal T} \tilde S^a_n {\cal T}^{-1} &=& \bar M_{a\dot b} \tilde S^{\dot b}_n ~.
\label{calT.act.gs}
\eea 
But notice that the states $\widetilde{|b \R}$ and the modes 
$\tilde S^{\dot b}_n$ are not defined in IIB. Although ${\cal T}$ does not 
have well defined action on the above objects in IIB, it does have well 
defined action on the states in (\ref{NSgs.basis}). This is because, each of 
these 
states is either a direct product of even number of spinor representations and 
the vector representation or odd number of spinor representations and the 
conjugate spinor representation. Therefore all of them tranform tensorially 
under $SO(8)$ and pick up some factor of ($-$) signs under ${\cal T}$ (since 
it 
is a reflection along certain directions). So to find the action of  ${\cal T}$
 on these states one needs to perform the tensorial decomposition of them. 
This, unfortunately is a difficult task. 

Thus we see that the problem arises because of the specific variables that we 
are working with or equivalently the specific choice of basis that we have 
made. It seems that it is best to come out of the language of this specific 
realization and adopt that of the curent algebra, which is universal, and then
reexpress these basis states in the GS formalism. 
\sectiono{Basis Construction and the Non-BPS Boundary States} 
\label{sec-basis}
\subsectiono{Basis Construction}
\label{subsec-basis}
An irreducible representation $\Pi^{(w)}$ is obtained by applying $H^i_{-n}$, 
$E^{\pm \alpha}_{-n}$ and $E^{-\alpha}_0$ on the corresponding highest weight 
state $|w\R$ in all possible ways, where $i=1,..,4$, $n>0$ and $\alpha \in 
\Delta^+$, $\Delta^+$ being the set of all positive roots. The Hilbert space 
contains null states, which one has to put to zero. Using eqns (\ref{HH}), 
(\ref{HE}), (\ref{EE}) and the fact that any $\alpha \in \Delta$ (set of all 
roots) can be written as a linear combination of the simple roots 
$\alpha^s_i$'s with integer coefficients, one can argue that it is sufficient 
to use $H^i_{-n}~(n>0)$ and $E^{\pm \alpha^s_i}_{-n}~(n\geq 0)$ to build up the
spectrum. But since $E^{\pm \alpha^s_i}_{-n}$'s $(n\geq 0)$ do not commute, 
using these is inconvenient. We solve this problem by treating the vacuum 
states (which are finite in number) in a special way and using the commuting 
modes $E^{\pm \alpha_i}_{-n}$'s ($n>0,i=1,...,4$) with $\alpha_i$'s defined in
eqns.~(\ref{roots}), instead of 
$E^{\pm \alpha^s_i}_{-n}$'s. Our basis states would look exactly similar to 
those one would have constructed in the bosonic fomalism. In this formalism,
a basis state is given certain compact momentum ($P$) and raised to some energy
 by applying the negative mode oscillators $\beta^i_{-n}$'s on the bosonic 
vacuum. We will use the $E^{\alpha_i}_{-n}$ oscillators to create the compact 
momentum (i.e to fix the position on the weight lattice) and use 
$H^i_{-n}(=\beta^i_{-n})$'s to
lift it to any desired energy. The construction will be free of null states. 

Since these basis states will be much like the bosonic states, it is useful to 
consider the bosonic expression for the zero-mode of the energy-momentum 
tensor:
\beq
L^{BOS}_0 = \frac{1}{2} P^2 + N^H ~,             \label{L0bos}
\eeq
where $P^2= \displaystyle{\sum_i P^i P^i}$, $P^i=H^i_0$ and  
$N^H= \displaystyle{\sum_i} \displaystyle{\sum_{n>o}} H^i_{-n} H^i_n$. We will 
call a state a 
bosonic vacuum state\footnote{Bosonic vacuum states should not be confused with
 the vacuum representation of the current algebra.}, if $N^H|state\R=0$. 
Therefore given the momentum $P$, $L^{BOS}_0$ takes its minimum value for the 
bosonic vacuum
state. While constructing the basis we will use the fact that any state in the 
 free module generated by $J^{\mu \nu}_{-n}$'s $(n\geq 0)$ having $L^{BOS}_0$
eigenvalue less than $\frac{1}{2} P^2$ eigenvalue, should be null. The same 
argument may be applied to show:
\beq
H^i_n |P,~N^H=0\R = 0~, \quad \quad \quad \forall n>0 ~, \label{bos.vac.state}
\eeq
where $|P,~N^H=0\R$ is a bosonic vacuum state.

Let us consider $\Pi^{(e)}$ first. The current algebra vacuum states, 
annihilated by $J^{\mu \nu}_n$, ($\forall \mu,\nu$ and $\forall n>0$) are 
$|\pm e_i\R$ (eqn.~(\ref{pmei})), where the states are labelled by their $P$
eigenvalue (see eqn.~(\ref{mom.ei})). If we adopt a new notation for these 
states, namely
$|l_1,l_2,l_3,l_4\R$ with $P = \sum_i l_i \alpha_i$, then we can write,
\bea
\begin{array}{ll}  
|\pm e_1\R = |\pm \frac{1}{2}, \pm \frac{1}{2},0,0 \R~, &
|\pm e_2\R = |\pm \frac{1}{2}, \mp \frac{1}{2},0,0 \R~, \\
 & \\
|\pm e_3\R = |0,0,\pm \frac{1}{2}, \pm\frac{1}{2} \R~,  & 
|\pm e_4\R = |0,0,\pm \frac{1}{2}, \mp\frac{1}{2} \R~.
\end{array}
\label{e.vac}
\eea
Now the allowed momenta in $\Pi^{(e)}$ take value in $\Lambda_e$.
We define a set ${\cal K}^{(e)}$ of 4-vectors $\vec s \equiv (s_1,s_2,s_3,
s_4)$, such that each $s_i$ $(i=1,..,4)$ is either integer or half-odd-integer
and,
\bea
\begin{array}{rcc}
\mbox{either} & \quad s_1, s_2 \in {\bf Z} + \frac{1}{2} ~, & 
s_3, s_4 \in {\bf Z} ~,\\
\mbox{or} & \quad s_1, s_2 \in {\bf Z} ~, & s_3, s_4 \in {\bf Z} + \frac{1}{2}
~.
\end{array} 
\label{def.Ke}
\eea
Then one can show that if $P=\sum_i s_i \alpha_i \in \Lambda_e$, then $\vec s
\in {\cal K}^{(e)}$ and vice versa. 
Notice that the $|\pm e_i\R$'s are examples of this. 
The bosonic vacuum state with momentum $P=\sum_i s_i \alpha_i 
\in \Lambda_e$ can be expressed as,
\bea
|\vec s \in {\cal K}^{(e)} \R = \prod_i {\cal E}^{\alpha_i}\lt(s_i \rt)
|V\R~,        \label{def.bos.vac1}
\eea
where the state $|V\R$ and the operators ${\cal E}^{\alpha_i}\lt(s_i \rt)$ are 
defined as follows:
\bea
|V\R \equiv \lt \{ \begin{array}{ll} 
|\pm e_1\R \quad & s_1,s_2  \in {\bf Z}^{\pm}_{\frac{1}{2}}~,\\
 & \\
|\pm e_2\R \quad & s_1 \in {\bf Z}^{\pm}_{\frac{1}{2}},~
s_2 \in {\bf Z}^{\mp}_{\frac{1}{2}}~,\\
&\\
|\pm e_3\R \quad & s_3,s_4  \in {\bf Z}^{\pm}_{\frac{1}{2}}~,\\
&\\
|\pm e_4\R \quad & s_3 \in {\bf Z}^{\pm}_{\frac{1}{2}},~
 s_4  \in {\bf Z}^{\mp}_{\frac{1}{2}} ~.
\end{array} \rt.   \label{def.bos.vac2}
\eea
\bea
{\cal E}^{\alpha}\lt(s \rt) \equiv \lt \{ \begin{array}{ll}
E^{\pm \alpha}_{-2|s|+1} E^{\pm \alpha}_{-2|s|+3}\cdots E^{\pm \alpha}_{-2} 
\quad & s \in {\bf Z}^{\pm}_{\frac{1}{2}}~, \\
&\\
E^{\pm \alpha}_{-2|s|+1} E^{\pm \alpha}_{-2|s|+3}\cdots E^{\pm \alpha}_{-1} 
\quad & s \in {\bf Z}^{\pm}~,\\
&\\
1 & s=0,~\pm \frac{1}{2}~,
\end{array} \rt.    \label{def.bos.vac3}  
\eea
where ${\bf Z}^{\pm}$ are respectively the sets of positive and negative 
integers and ${\bf Z}^{\pm}_{\frac{1}{2}}$ are respectively the sets of 
positive and negative half-odd-integers. As all $E^{\alpha_i}$'s are commuting,
 the ordering of the operators 
${\cal E}^{\alpha_i}\lt(s_i \rt)$ in the product in eqn.~(\ref{def.bos.vac1})
is not important. 

The fact that the state $|\vec s\R$ defined above, is a bosonic vacuum state 
can be proved as follows: using eqn.~(\ref{HE}) for $m=0$ and 
eqn.~(\ref{mom.ei}), it can be shown that,
\bea 
P|\vec s\R = \vec s~ |\vec s\R~.
\eea
{}From the Sugawara's construction of the energy-momentum tensor 
(eqns.~(\ref{Tsug}), (\ref{JJ.conf.ord})), the following commutation relation
can be established:
\bea
\lt[ L_0^{SUG}, J_n^{\mu \nu}\rt] = - n J^{\mu \nu}_n~, \quad \quad 
\forall n\in {\bf Z}~.    \label{L0Jn}
\eea
The $L_0^{SUG}$ eigenvalues of the vacuum states can also be found:
\bea
L_0^{SUG} |\pm e_i\R = \frac{1}{2} P^2~|\pm e_i\R = \frac{1}{2}~|\pm e_i\R~.
\label{pmei.e.value}
\eea
Then using eqns.~(\ref{L0Jn}) and (\ref{pmei.e.value}), one can show that, 
\bea
L_0^{SUG} |\vec s\R = \frac{1}{2} P^2~|\vec s\R = \lt( \sum_i s_i^2 \rt) 
|\vec s\R~.
\eea
Hemce $|\vec s\R$ is a bosonic vacuum state.

In Appendix \ref{app-basis} we show that the norm of a bosonic vacuum
 state constructed above is 1. The orthogonality of the basis states can be 
shown as follows: for a given momentum eigenvalue corresponding to a point on 
the lattice $\Lambda_e$, there is only one linearly independent state which 
qualifies as a bosonic vacuum state (this fact is manifest in the bosonic 
formalism). Now since $H^i_0 = P^i$ is a hermitian operator, states with 
different momentum eigenvalues must be orthogonal.    

Finally the full set of orthonormal basis in $\Pi^{(e)}$ is given by 
$\lt\{ \left|\vec s\in {\cal K}^{(e)},~\lt\{N_{in} \rt\} \right\R \rt\}$,
\bea
\left|\vec s \in {\cal K}^{(e)},~\lt\{N_{in} \rt\} \right\R  \equiv 
\prod_{\stackrel{n>0,}{i}}\frac{ \lt( \frac{1}{\sqrt n} H^i_{-n} 
\rt)^{N_{in}}}
{\sqrt{N_{in}!}} |\vec s \in {\cal K}^{(e)} \R~,
\label{def.basis.e}
\eea
with $N_{in} = 0,1,2,\cdots ~\forall i$, $\forall n>0$. The following 
orthonormality condition of this complete set of basis states can be checked by
using eqn.~(\ref{HH}) and the orthonormality of the bosonic vacuum states:
\bea
\L \vec s^{~\prime} \in {\cal K}^{(e)}, \{N_{in}^{\prime}\}|
\vec s \in {\cal K}^{(e)}, \{N_{in}\}\R = \delta_{{\vec s}^{\,\prime},\vec s}~~
\delta_{\{N_{in}^{\prime}\},\{N_{in}\}}
\eea

The right moving states are defined, as before, by replacing the left handed 
oscillators by the right handed ones. The same procedure can be followed for 
$\Pi^{(\bar \delta)}$ and $\Pi^{(\delta)}$. But we do not need the explicit 
expressions for these two, as we are interested in writing down the non-BPS 
states which involve only $\Pi^{(e)}$.  
\subsectiono{Non-BPS Boundary States}
\label{subsec-nbps}
We have seen in subsec.~\ref{subsec-IIBbranes} that the actual non-BPS 
boundary state takes the form $\bar{\cal A}|p,e~\R\R$, where $\bar{\cal A}$
is an operator which does not involve fermionic oscillators and hence takes 
the same form in both NSR and GS formalisms. Therefore it is sufficient to 
consider only the Ishibashi state corresponding to the highest weight $w=e$ 
for a given non-BPS alignment and try to express that in terms of the current
modes. Let us start with the following Ishibashi state:  
\bea
|{\cal N},e~\R\R = \sum_{\vec s\in {\cal K}^{(e)}, \lt\{N_{in}\rt\}} 
|\vec s,\lt\{N_{in} \rt\} \R
 \otimes {\cal T}_{\cal N} \Theta 
\widetilde{\left|\vec s,\lt\{N_{in} \rt\} \right\R} ~,    
\label{Ishib.1}
\eea
where ${\cal T}_{\cal N}$ is the Hilbert space operator (acting only on the 
right part) corresponding to the reflection along the following set of ordered
directions: 
\beq
{\cal N} = \lt \{ \mu_1, \mu_2,\cdots ,\mu_{p+1} \rt \}_{\mu_1<\mu_2<\cdots 
<\mu_{p+1}}~.
\eeq
In the previous subsection the basis states have been written purely in terms 
of the currents. Once we get the action of ${\cal T}_{\cal N} \Theta$ on these
and finally express everything in terms of the current modes, the above 
Ishibashi state can be translated into the language of any of the three 
realizations considered in sec.~\ref{sec-realization}. Since the above state 
involves only the NS-NS sector exitations, this is realized both in IIA and 
IIB. When realized in IIA with $p$=odd, this state gives a non-BPS state in 
IIA. Similarly in IIB, this gives a non-BPS state for $p$=even. 

Let us now analyze the action of $\Theta$ on the basis states. Using 
eqns.~(\ref{theta.act}), (\ref{HJ}), (\ref{EJ}), (\ref{pe1.rep.}) and 
(\ref{me1.rep.}) and the fact that $\Theta$ is antilinear, 
we get the following:
\bea
\begin{array}{rll}
\Theta \tilde H^i_n \Theta^{-1} & = - \tilde H^i_n~, \quad
&\forall i, \forall n>0~, \\
\Theta \tilde E^{\alpha_i}_n \Theta^{-1} & = - \tilde E^{-\alpha_i}_n~, \quad
&\forall i, \forall n>0~, \\
\Theta \widetilde{|\pm e_i\R} & = \widetilde{|\mp e_i\R}~, \quad &\forall i ~. 
\end{array}  \label{theta.act.HEpme}
\eea
Now applying the second and third eqns. of (\ref{theta.act.HEpme}) and 
eqns.~(\ref{def.bos.vac1}), (\ref{def.bos.vac2}), (\ref{def.bos.vac3}) one can 
show:
\bea
\Theta \widetilde{|\vec s \R} =  (-1)^{\scriptstyle \displaystyle \sum_i 
|s_i|-1} \widetilde{|-\vec s \R}~\equiv~ |\overline{ -\vec s}\R ~, 
\label{theta.act.bos.vac}
\eea
where $\vec s \in {\cal K}^{(e)}$. Then finally using the 
eqns.~(\ref{Ishib.1}), (\ref{def.basis.e}), (\ref{theta.act.bos.vac}) and the 
first eqn. of (\ref{theta.act.HEpme}) one can show:
\bea
|{\cal N},e~\R\R = {\cal T}_{\cal N}~ \exp\lt[- \sum_{\stackrel{n>0,}{i}} 
\frac{1}{n} H^i_{-n} \tilde H^i_{-n} \rt] 
\sum_{\vec s\in {\cal K}^{(e)}} |\vec s \R \otimes |\overline{-\vec s }\R~.
\label{Ishib.2}
\eea
Now let us define a 4-vector $\vec n\lt(\vec s\rt)=\lt(n_1(\vec s),n_2(\vec 
s),n_3(\vec s),n_4(\vec s)\rt)$ by,
\bea
\sum_i n_i\lt(\vec s\rt) e_i = \sum_i s_i \alpha_i~.
\eea
Then one can show that $\forall \vec s\in {\cal K}^{(e)}$, $\exists$ a unique 
$\vec n(\vec s)$ such that $n_i(\vec s) \in {\bf Z}$, and 
$\sum_i n_i(\vec s) = \hbox{odd}$ and vice versa. Therefore there is a 
one-to-one correspondence between the vectors $\vec s$ and $\vec n$. 
We will adopt the following notation for the Ishibashi state (\ref{Ishib.2}) 
in future:
\bea
|{\cal N},e~\R\R = {\cal T}_{\cal N}~ \exp\lt[- \sum_{\stackrel{n>0,}{i}} 
\frac{1}{n} H^i_{-n} \tilde H^i_{-n} \rt] 
\sum_{\vec n}^{odd} |\vec s(\vec n) \R \otimes |\overline{-\vec s(\vec n) }\R~,
\label{Ishib.3}
\eea
where $\displaystyle{\sum_{\vec n}^{odd}}$ represents sum over integer $n_i$'s 
with 
$\displaystyle{\sum_i n_i}=\mbox{odd} $ and $\vec s(\vec n)$ is the inverse 
function of
$\vec n(\vec s)$. For ${\cal N}=\{\hbox{null}\}$, i.e., ${\cal T}_{\cal N}=$ 
identity, this state, expressed in terms of IIA variables, gives the non-BPS 
D-instanton in IIA (also this same state, expressed in terms of the IIB 
variables, gives the NS-NS part of the BPS D-instanton boundary state in IIB). 
Let us now consider a nontrivial case where ${\cal T}_{\cal N}$ is not 
identity.\\

\noindent {\bf Reflection along the direction $\mu$ =2:} In this case 
${\cal N}=\{2\}$. Equation (\ref{calT.act}) reads:
\bea
{\cal T}_{\{2\}} \tilde J^{\mu \nu}_n {\cal T}_{\{2\}}^{-1} = \lt \{
\begin{array}{ll}
\tilde J^{\mu \nu}_n~, \quad & \mu \neq 2,~ \nu \neq 2 ~,\\
-\tilde J^{\mu \nu}_n~, \quad & \hbox{otherwise}~.
\end{array} \rt.
\eea
Then using eqns.~(\ref{HJ}) and (\ref{EJ}) one shows,
\bea
{\cal T}_{\{2\}} \tilde H^i_n {\cal T}_{\{2\}}^{-1} = 
\lt \{ \begin{array}{ll}
- \tilde H^i_n~, \quad & i=1 ~,\\
\tilde H^i_n~, \quad & i\neq 1 ~.
\end{array} \rt.     \label{T2act.H} 
\eea
\bea
{\cal T}_{\{2\}} \tilde E^{\pm \alpha_1}_n {\cal T}_{\{2\}}^{-1}
= \tilde E^{\mp \alpha_2}_n~,&&
{\cal T}_{\{2\}} \tilde E^{\pm \alpha_2}_n {\cal T}_{\{2\}}^{-1}
= \tilde E^{\mp \alpha_1}_n~, \nonumber\\
{\cal T}_{\{2\}} \tilde E^{\pm \alpha_3}_n {\cal T}_{\{2\}}^{-1}
= \tilde E^{\pm \alpha_3}_n~,&&
{\cal T}_{\{2\}} \tilde E^{\pm \alpha_4}_n {\cal T}_{\{2\}}^{-1}
= \tilde E^{\pm \alpha_4}_n~.       \label{T2act.E}
\eea
Action of ${\cal T}_{\{2\}}$ on the current algebra vacuum states in the 
covariant basis is as follows:
\bea
{\cal T}_{\{2\}} \widetilde{|\mu \R} = \lt\{ \begin{array}{rl}
-\widetilde{|\mu \R}~, \quad \quad & \mu=2~, \\
\widetilde{|\mu \R}~, \quad \quad & \mu \neq 2~.
\end{array} \rt.         \label{T2act.mu}
\eea
Using this and eqn.~(\ref{pmei}) one can show:
\bea
{\cal T}_{\{2\}} \widetilde{|\pm e_1 \R} &=& \widetilde{|\mp e_1 \R} ~,
\nonumber\\
{\cal T}_{\{2\}} \widetilde{|\pm e_i \R} &=& \widetilde{|\pm e_i \R}~, 
\quad \quad i\neq1~.
\label{T2act.pmei}
\eea
One can check that eqns.~(\ref{T2act.pmei}) are consistent with the 
transformations (\ref{T2act.E}) and the eqns.~(\ref{pe1.rep.}), 
(\ref{me1.rep.}). Now using the transformations (\ref{T2act.E}) and 
(\ref{T2act.pmei}), one can show that,
\bea
{\cal T}_{\{2\}} |\vec s(\vec n) \R \otimes |\overline{-\vec s(\vec n) }\R = 
|\vec s(\vec n) \R \otimes |\overline{- \vec s\lt(D_{\{2\}}\vec n \rt)}\R ~,  
\label{T2act.bos.vac}
\eea
where $D_{\{2\}} = \hbox{diag} (-1,1,1,1)$. Finally applying 
eqns.~(\ref{Ishib.3}), (\ref{T2act.H}) and (\ref{T2act.bos.vac}) one gets,
\bea
|\{2\},e~\R\R &=& {\cal T}_{\{2\}}~ \exp\lt[- \sum_{\stackrel{n>0,}{i}} 
\frac{1}{n} H^i_{-n} \tilde H^i_{-n} \rt] 
\sum_{\vec n}^{odd} |\vec s(\vec n) \R \otimes |\overline{-\vec s(\vec n) }\R~,
\nonumber\\
&=& \exp\lt[- \sum_{n>0} \frac{1}{n} H^{T}_{-n} D_{\{2\}} \tilde H_{-n} \rt] 
\sum_{\vec n}^{odd} 
|\vec s(\vec n) \R \otimes |\overline{- \vec s\lt(D_{\{2\}}\vec n \rt)}\R~.
\quad \quad
\label{Ishb.4}
\eea

We will give explicit expressions like eqn.~(\ref{Ishb.4}) for the Ishibashi 
states for some special alignments first. We define these special alignments 
as follows:
\bea
\begin{array}{llll}
{\cal N}^A_1=\{null\}, & {\cal N}^A_2=\{2,4\}, & {\cal N}^A_3=\{2,6\}, &
{\cal N}^A_4=\{2,8\}, \\
{\cal N}^A_5=\{4,6\}, & {\cal N}^A_6=\{4,8\}, &{\cal N}^A_7=\{6,8\}, &
{\cal N}^A_8=\{2,4,6,8\}.
\end{array}       \label{NA}
\eea
\bea
\begin{array}{llll}
{\cal N}^B_1=\{2\}, & {\cal N}^B_2=\{4\}, & {\cal N}^B_3=\{6\}, &
{\cal N}^B_4=\{8\}, \\
{\cal N}^B_5=\{2,4,6\}, & {\cal N}^B_6=\{2,4,8\}, & {\cal N}^B_7=\{2,6,8\}, & 
{\cal N}^B_8=\{4,6,8\}. 
\end{array}        \label{NB}
\eea
Every special alignment ${\cal N}^A_{\rho}$ $(\rho = 1,2,...,8)$ corresponds
to some non-BPS brane in IIA. Similarly every ${\cal N}^B_{\rho}$ corresponds 
to some non-BPS brane in IIB. To get the Ishibashi state for any of these 
special alignments one follows the same procedure as that we have just 
described for ${\cal N}^B_1$ (eqn.~(\ref{Ishb.4})). We give the results below:
\bea
|{\cal N}^A_{\rho},e~\R\R = \exp\lt[ -\sum_{n>0} \frac{1}{n} 
H^{T}_{-n} D^A_{\rho} \tilde H_{-n} \rt] 
\sum_{\vec n}^{odd} 
|\vec s(\vec n) \R \otimes |\overline{- \vec s\lt(D^A_{\rho}\vec n \rt)}\R~,
\label{Ishib.NA}
\eea
\bea
|{\cal N}^B_{\rho},e~\R\R = exp\lt[ -\sum_{n>0} \frac{1}{n} 
H^{T}_{-n} D^B_{\rho} \tilde H_{-n} \rt] 
\sum_{\vec n}^{odd} 
|\vec s(\vec n) \R \otimes |\overline{- \vec s\lt(D^B_{\rho}\vec n \rt)}\R~,
\label{Ishib.NB}
\eea
where the $D$ matrices are given in Table \ref{tab.D}. 
\begin{table}[h]
\begin{center}
\begin{tabular}{|c|c|c|} \hline\hline
$ \rho $ & $ D^A_{\rho} $           & $ D^B_{\rho} $      \\ \hline\hline
$1$      & ${\bf 1}_4$              & diag $(-1,1,1,1)$   \\ \hline
$2$      & diag$(-1,-1,1,1)$        & diag$(1,-1,1,1)$    \\ \hline
$3$      & diag$(-1,1,-1,1)$        & diag$(1,1,-1,1)$    \\ \hline
$4$      & diag$(-1,1,1,-1)$        & diag$(1,1,1,-1)$    \\ \hline
$5$      & diag$(1,-1,-1,1)$        & diag$(-1,-1,-1,1)$  \\ \hline
$6$      & diag$(1,-1,1,-1)$        & diag$(-1,-1,1,-1)$  \\ \hline
$7$      & diag$(1,1,-1,-1)$        & diag$(-1,1,-1,-1)$  \\ \hline
$8$      & $-{\bf 1}_4$             & diag $(1,-1,-1,-1)$ \\ \hline
\end{tabular}
\end{center}
\caption{The $D$ matrices} \label{tab.D}
\end{table}

Since the final aim is to get the non-BPS boundary state in the GS formalism, 
we mention here precisely how one translates the above state in terms of the 
GS variables. From eqns.~(\ref{Jgs}) and the mode expansions 
(\ref{JHEcurrents}) we get the following:
\bea
\begin{array}{lll}
J^{\mu \nu}_n =& \frac{i}{4} \lt(\gamma^{\mu} {\bar \gamma}^{\nu} \rt)_{ab}
\displaystyle{\sum_{m\in {\bf Z}}} 
{\stackrel{\scriptstyle \circ} {\scriptstyle \circ}}
S^a_m S^b_{n-m}{\stackrel{\scriptstyle \circ} {\scriptstyle \circ}}
 \quad & \lt(\hbox{IIB, IIA}\rt)~, \\
&&\\
\tilde J^{\mu \nu}_n =& \frac{i}{4} \lt({\gamma}^{\mu} \bar \gamma^{\nu}
\rt)_{ab} \displaystyle{\sum_{m\in {\bf Z}}} 
{\stackrel{\scriptstyle \circ} {\scriptstyle \circ}}
\tilde S^{a}_m \tilde S^{b}_{n-m}
{\stackrel{\scriptstyle \circ} {\scriptstyle \circ}}
 \quad & \lt(\hbox{IIB}\rt)~,\\
&&\\
\tilde J^{\mu \nu}_n =& \frac{i}{4} \lt({\bar \gamma}^{\mu} \gamma^{\nu}
\rt)_{\dot a\dot b} \displaystyle{\sum_{m\in {\bf Z}}} 
{\stackrel{\scriptstyle \circ} {\scriptstyle \circ}}
\tilde S^{\dot a}_m \tilde S^{\dot b}_{n-m}
{\stackrel{\scriptstyle \circ} {\scriptstyle \circ}}
\quad & \lt(\hbox{IIA}\rt)~.
\end{array} 
\label{Jgs.modes}
\eea
The oscillator normal ordering used above is defined in the following way:
\bea
{\stackrel{\scriptstyle \circ} {\scriptstyle \circ}}
S^a_m S^b_n {\stackrel{\scriptstyle \circ} {\scriptstyle \circ}}
&=& \lt\{\begin{array}{rl}
S^a_m S^b_n & \quad m \leq 0~, \\
-S^b_n S^a_m & \quad m > 0~.
\end{array} \rt. \label{osc.norm.ord}
\eea
Then
one uses eqns.~(\ref{HJ}), (\ref{EJ}) and (\ref{Jgs.modes}) to translate the 
$H^i_{-n}$, $E^{\alpha_i}_{-n}$ ($i=1,\cdots ,4$, $n>0$) oscillators in terms 
of the GS variables. Since all the states in the expansion of the Ishibashi 
states (\ref{Ishib.NA}) and (\ref{Ishib.NB}) are defined in terms of these 
current modes
 and the current algebra vacuum states\footnote{These states in the lattice 
basis are defined in eqn.~(\ref{pmei}) in terms of the covariant basis vacuum 
states $|\mu \R$, which are in turn the same as the states in the 
representation of the fermionic zero mode oscillators (see the last paragraph 
of sec.~\ref{sec-realization}).} $|\pm e_i\R$, using these one can translate 
the Ishibashi states in terms of the GS oscillators.

Let us now explain why we considered the ``special alignments'' to write 
down the explicit form of the Ishibashi states and in what sense they are 
special. In a formalism
with explicit covariance, no direction is special unless it is physically 
distinct. In the present case, it is the noncovariant nature of the basis 
construction\footnote {It has been mentioned in sec.~\ref{sec-int} that to 
make the basis, constructed in terms of currents, free of null states, one has 
to break the explicit covariance.} which is responsible for this. As a
result the Ishibashi states corresponding to the same world-volume dimension, 
but with different alignments may look very different, although they can be 
connected by some $SO(8)$ rotation. This is simply because the ${\cal T}$
operators for these two cases act very differently on the basis states due to
the non-covariance. Let us consider the special subset of boundary conditions 
(\ref{J.b.c}) for which the CSA oscillators ($H^i_n$'s) and the nonzero root 
oscillators ($E^{\alpha}_n$) do not get mixed up by ${\cal T}$ 
(eqn.~(\ref{calT.act})). We will call a boundary condition of this type a CSA 
preserving boundary condition and the corresponding alignment a CSA preserving
alignment. The special alignments correspond to boundary conditions of
 this type. Furthermore, the Ishibashi state for any CSA preserving alignment 
${\cal C}$ can always be obtained from one of the specially aligned Ishibashi 
states by applying a rotation operator (which depends on ${\cal C}$), generated
 by $\tilde H^i_0$'s or $\lt(H^i_0 + \tilde H^i_0 \rt)$'s, on it. Such a 
rotation 
operator goes through the exponential operator in the eqn.~(\ref{Ishib.NA}) or 
(\ref{Ishib.NB}) and acts on the bosonic vacuum states to give a phase 
$\varsigma^{({\cal C})} \lt( \vec s(\vec n) \rt)$. Therefore the general 
structure of an Ishibashi state with alignment ${\cal C}$, is the following:
\bea
|{\cal C}, e~\R\R = \exp\lt[ -\sum_{n>0} \frac{1}{n} 
H^{T}_{-n} D \tilde H_{-n} \rt] \sum_{\vec n}^{odd} 
\varsigma^{({\cal C})} \lt( \vec s(\vec n) \rt) |\vec s(\vec n) \R \otimes 
|\overline{- \vec s\lt(D \vec n \rt)}\R~,
\label{Ishib.C}
\eea
where $D$ is one of the matrices given in Table 1. This state can be rewritten 
in terms of the GS variables in the same way as discussed before.

The rotation operators required for generating all the CSA preserving 
alignments from the special alignments given in eqns.~(\ref{NA}), (\ref{NB})
are, 
\bea
{\cal U}_i &\equiv& e^{i \pi \tilde H^i_0}~, \nonumber\\
{\cal R}_i \lt( \theta_i \rt) &\equiv& e^{i \theta_i \lt(H^i_0 + \tilde H^i_0
\rt) }~.
\eea
Notice that ${\cal R}_i \lt( \theta_i \rt)$ acts both on the right and left 
parts and hence gives a physical rotation to a brane by an angle $\theta_i$ 
on the $(2i-1)$-$2i$ plane. But since ${\cal U}_i$ acts only on the right part,
generically, acting on some state this gives another physically inequivalent 
state (state which is not connected to the initial one through any symmetry 
transformation) in the Hilbert space\footnote{There are exceptions which we are
 going to discuss below.}. ${\cal U}_i$ provides the reflection on the 
coordinates 
$\mu = 2i-1,~2i$. Acting on an Ishibashi state which is not aligned along the 
$(2i-1)$-$2i$ plane, ${\cal U}_i$ extends its world-volume to this additional 
plane and similarly acting on an Ishibashi state already aligned along the 
$(2i-1)$-$2i$ plane, it reduces its world-volume by this plane. If the initial 
Ishibashi state has one of the two directions 
$\mu=2i-1, 2i$ to be tangential and the other normal, then ${\cal U}_i$ 
interchanges these tangential and normal directions. But this is same as 
performing a physical rotation on the $(2i-1)$-$2i$ plane through $\pi/2$ which
could also be achieved by applying ${\cal R}_i \lt( \pi/2\rt)$. Given all 
these, following will be the prescription for generating the whole class of 
Ishibashi states starting from that of the corresponding special alignment:
Take the Ishibashi state of the special alignment and operate only those 
${\cal U}_i$ and ${\cal R}_i \lt( \theta_i \rt)$ operators on it to get the 
whole class of Ishibashi states, which are enlisted in Tables 2 and 3. The 
${\cal U}_i$ operators which are excluded from 
the list give over-counting, because the same state is obtained by applying 
${\cal R}_i \lt( \pi/2\rt)$. The ${\cal R}_i \lt( \theta_i \rt)$ operators that
 have been excluded from the list have trivial action on the initial state, 
because they provide rotations on the planes which are either orthogonal or 
tangential to the initial state.  
   
Now let us discuss the other alignments which do not preserve the CSA space. 
An Ishibashi state corresponding to this type of boundary condition will look
considerably different from those discussed above. In this case, one again 
starts from the eqn. similar to (\ref{Ishib.3}), but now with the following 
general form of ${\cal T}$: 
\bea
{\cal T} = \exp\lt[ i f\lt(H^i_0,~E^{\alpha}_0\rt)\rt]~,
\eea
where $f\lt(H^i_0,~E^{\alpha}_0\rt)$ is a linear function of its arguments such
 that ${\cal T}$ is hermitian (because ${\cal T}$ is unitary and also squares 
to
identity in the vector representation $\Pi^{(e)}$). We can treat this type
of Ishibashi states in the following way: Any Ishibashi state $|f,e~\R\R$
with CSA non-preserving boundary condition corresponding to the function
$f\lt(H^i_0,~E^{\alpha}_0\rt)$, can be written in the following form:        
\bea
|f,e~\R\R = {\cal O} \lt(f,{\cal C} \rt) |{\cal C},e~\R\R~, 
\eea
where the state $|{\cal C},e~\R\R$ is defined in eqn.~(\ref{Ishib.C}) and 
 ${\cal O} \lt(f,{\cal C} \rt)$ is an $SO(8)$ rotation operator which acts both
 on the right and the left parts. For a given function 
$f\lt(H^i_0,~E^{\alpha}_0\rt)$, the rotation operator ${\cal O}$ depends on the
 specific CSA preserving alignment ${\cal C}$ chosen on the right hand side of 
the above eqn. To get the state $|f,e~\R\R$ in GS formalism one can first 
translate the state $|{\cal C},e~\R\R$ in terms of the GS variables 
(following the procedure that has already been discussed), then apply the known
 rotation operator ${\cal O} \lt(f,{\cal C} \rt)$ (under which the GS variables
transform covariantly) on that.       
\begin{table}[p]
\begin{center}
\begin{tabular}{|c|c|c|c|} \hline\hline
$\rho$&$N^A_{\rho}$  &${\cal U}_i$&${\cal R}_i\lt(\theta_i\rt)$\\ \cline{3-4} 
      &              &    $i$     &            $i$             \\ \hline\hline
$1$   &$\{null\}$    &  1,2,3,4   &            --              \\ \hline
$2$   &$\{2,4\}$     &    3,4     &            1,2             \\ \hline
$3$   &$\{2,6\}$     &    2,4     &            1,3             \\ \hline
$4$   &$\{2,8\}$     &    2,3     &            1,4             \\ \hline
$5$   &$\{4,6\}$     &    1,4     &            2,3             \\ \hline
$6$   &$\{4,8\}$     &    1,3     &            2,4             \\ \hline
$7$   &$\{6,8\}$     &    1,2     &            3,4             \\ \hline
$8$   &$\{2,4,6,8\}$ &    --      &          1,2,3,4           \\ \hline
\end{tabular}
\end{center}
\caption{Allowed ${\cal U}_i$ and ${\cal R}_i\lt(\theta_i\rt)$
operators for $\lt\{N^A_{\rho}\rt\}$~.}   \label{tab.TR.NA} 
\end{table}
\begin{table}[p]
\begin{center}
\begin{tabular}{|c|c|c|c|} \hline\hline
$\rho$&$N^B_{\rho}$  &${\cal U}_i$&${\cal R}_i\lt(\theta_i\rt)$\\ \cline{3-4} 
      &              &    $i$     &            $i$             \\ \hline\hline
$1$   &$\{2\}$       &   2,3,4    &             1              \\ \hline
$2$   &$\{4\}$       &   1,3,4    &             2              \\ \hline
$3$   &$\{6\}$       &   1,2,4    &             3              \\ \hline
$4$   &$\{8\}$       &   1,2,3    &             4              \\ \hline
$5$   &$\{2,4,6\}$   &     4      &           1,2,3            \\ \hline
$6$   &$\{2,4,8\}$   &     3      &           1,2,4            \\ \hline
$7$   &$\{2,6,8\}$   &     2      &           1,3,4            \\ \hline
$8$   &$\{4,6,8\}$   &     1      &           2,3,4            \\ \hline
\end{tabular}
\end{center}
\caption{Allowed ${\cal U}_i$ and ${\cal R}_i\lt(\theta_i\rt)$
operators for $\lt\{N^B_{\rho}\rt\}$~.}   \label{tab.TR.NB}
\end{table}
\newpage
\appendix
\sectiono{Vacuum Representations}
\label{app-vac.rep}
\setcounter{equation}{0}
Here we will give the explicit vacuum representations $\Pi^{(w)}_0$ for 
$w = 0,~e,~\bar{\delta},~\delta$.
\begin{enumerate}
\item
{\bf $\Pi^{(0)}_0$:}

This contains only one state $|0\R$ defined by:
\beq
L^{SUG}_0 |0\R = J^{\mu \nu}_n |0\R = 0~, \quad \quad \forall n
\geq 0 ~. \label{sl2cvac}
\eeq
This is the $SL(2,{\bf C})$ and $SO(8)$ invariant vacuum state.\\

\item
{\bf $\Pi^{(e)}_0$:}

The orthonormal states are denoted $|\mu\R,~~\mu = 1,2,...,8$ and the 
representation is given by (see eqn.~(\ref{def.vac.repns.})):
\beq
T^{\mu \nu}_{(e)} = i \Sigma^{\mu \nu}  \label{Te}~,
\eeq
where $\Sigma^{\mu \nu}$ is an $8\times 8$ matrix, elements of which are given
 by:
\beq
\lt(\Sigma^{\mu \nu}\rt)_{\rho \sigma} = \delta^{\mu \rho}  \delta^{\nu \sigma}
- \delta^{\mu \sigma} \delta^{\nu \rho}~.      \label{sigma}
\eeq 
Now define states in the lattice basis as,
\beq
|\pm e_j\R \equiv \frac{1}{\sqrt{2}} \lt( |\mu =2j-1\R  \mp i|\mu =2j\R \rt)~.
\label{pmei}
\eeq
{}From the representation of $J_0^{\mu \nu}$ given above 
(eqns.~(\ref{def.vac.repns.}) ,(\ref{Te})), one can find the representations 
for the 
$H_0^i$'s in $\Pi^{(e)}_0$ by using eqn.~(\ref{HJ}). From this the $H_0^i=P^i$ 
eigenvalues of the states $|\pm e_j\R$ can be computed. One finds,
\bea
P |\pm e_i\R = \pm e_i~ |\pm e_i\R~.    \label{mom.ei}
\eea
where $e_i$'s are weight vectors defined in eqns.~(\ref{e}).
Using eqn.~$(\ref{EJ})$, the matrices for $E^{\alpha}_0$ can also be computed.
One can show the following relations:
\bea
|-e_1\R = E^{-\alpha_1}_0 E^{-\alpha_2}_0 |e_1\R~,~~~~~~~~~~~~~~~~~~~~~~~\,\cr
\begin{array}{ll}
|e_2\R =iE^{-\alpha_2}_0|e_1\R~,& \quad |-e_2\R =iE^{-\alpha_1}_0 |e_1\R~,\\
|e_3\R =iE^{-\alpha_6}_0|e_1\R~,& \quad |-e_3\R =iE^{-\alpha_5}_0 |e_1\R~,\\
|e_4\R =iE^{-\alpha_8}_0|e_1\R~,& \quad |-e_4\R =iE^{-\alpha_7}_0 |e_1\R~.
\end{array}
\label{pe1.rep.}
\eea
\bea
|e_1\R = E^{\alpha_1}_0 E^{\alpha_2}_0 |-e_1\R~,~~~~~~~~~~~~~~~~~~~~~~~~~~~~ \cr
\begin{array}{ll}
|e_2\R =iE^{\alpha_1}_0|-e_1\R~, & |-e_2\R =iE^{\alpha_2}_0 |-e_1\R~,\\
|e_3\R =iE^{\alpha_5}_0|-e_1\R~, & |-e_3\R =iE^{\alpha_6}_0 |-e_1\R~,\\
|e_4\R =iE^{\alpha_7}_0|-e_1\R~, & |-e_4\R =iE^{\alpha_8}_0 |-e_1\R~.
\end{array}
\label{me1.rep.}
\eea
\\
\item
{\bf $\Pi^{(\bar{\delta})}_0$:}

The orthonormal states are denoted $|\dot{a}\R,~~\dot{a}=1,2,...,8$ and the 
representation is given by:
\beq
T^{\mu \nu}_{(\bar{\delta})} = \frac{i}{2} \bar{\gamma}^{\mu} \gamma^{\nu}~.
  \label{Tdb}
\eeq 
(For definition and explicit reprsentation of the $8\times 8$ matrices 
$\gamma^{\mu}$ and  $\bar{\gamma}^{\mu}$, see appendix \ref{app-gamma}.) The 
states in the lattice basis are:
\beq
|\pm \bar{\delta}_j\R \equiv \frac{1}{\sqrt{2}} \lt( |\dot{a} =2j-1\R  \mp i
|\dot{a} =2j\R \rt)~,
\label{pmdbi}
\eeq
where $\bar{\delta}_j$'s are the following four conjugate spinor weights:
\beq
\bar{\delta}_1 = \frac{1}{2}\pmatrix{1\cr1\cr-1\cr1}, 
\bar{\delta}_2 = \frac{1}{2}\pmatrix{1\cr1\cr1\cr-1},
\bar{\delta}_3 = \frac{1}{2}\pmatrix{-1\cr1\cr1\cr1}, 
\bar{\delta}_4 = \frac{1}{2}\pmatrix{-1\cr1\cr-1\cr-1}.      \label{db}
\eeq
Following the same procedure as described for eqn.~(\ref{mom.ei}) one can 
establish the following eigenvalue equation for the present representation:
\bea
P |\pm \bar \delta_i \R = \pm \bar \delta_i~ |\pm \bar \delta_i \R~. 
\label{mon.deltai.bar}
\eea
Similarly as before one can check the following relations:
\bea
\begin{array}{ll}
|-\bar{\delta}_1\R = -iE^{-\alpha_1}_0 |\bar{\delta}_2\R~, &\quad 
|\bar{\delta}_1\R = E^{-\alpha_4}_0|\bar{\delta}_2\R~, \\
|-\bar{\delta}_2\R = iE^{-\alpha_1}_0 E^{-\alpha_4}_0 |\bar{\delta}_2\R~,&\\ 
|-\bar{\delta}_3\R = iE^{-\alpha_9}_0 |\bar{\delta}_2\R~, &\quad
|\bar{\delta}_3\R = -E^{-\alpha_8}_0 |\bar{\delta}_2\R~, \\
|-\bar{\delta}_4\R = -E^{-\alpha_{12}}_0 |\bar{\delta}_2\R~, &\quad
|\bar{\delta}_4\R = -iE^{-\alpha_5}_0|\bar{\delta}_2\R~.
\end{array}
\label{pdb2.rep.} 
\eea
\bea
\begin{array}{ll}
|\bar{\delta}_1\R = -iE^{\alpha_1}_0|-\bar{\delta}_2\R~,& \quad
|-\bar{\delta}_1\R = -E^{\alpha_4}_0 |-\bar{\delta}_2\R~, \\
|\bar{\delta}_2\R = -i E^{\alpha_1}_0 E^{\alpha_4}_0 |-\bar{\delta}_2\R~, &\\
|\bar{\delta}_3\R = iE^{\alpha_9}_0|-\bar{\delta}_2\R~,& \quad
|-\bar{\delta}_3\R = E^{\alpha_8}_0 |-\bar{\delta}_2\R~, \\
|\bar{\delta}_4\R = E^{\alpha_{12}}_0|-\bar{\delta}_2\R~, & \quad
|-\bar{\delta}_4\R = -iE^{\alpha_5}_0 |-\bar{\delta}_2\R~. 
\end{array}
\label{mdb2.rep.}
\eea
\\
\item
{\bf $\Pi^{(\delta)}_0$:}

The orthonormal states are denoted $|a\R,~~a=1,2,...,8$ and the representation 
is given by:
\beq
T^{\mu \nu}_{(\delta)} = \frac{i}{2} \gamma^{\mu} \bar{\gamma}^{\nu}~.
  \label{Td}
\eeq 
The states in the lattice basis are:
\beq
|\pm \delta_j\R \equiv \frac{1}{\sqrt{2}} \lt( |a =2j-1\R  \mp i|a =2j\R \rt)~,
\label{pmdi}
\eeq
where $\delta_j$'s are the following four spinor weights:
\beq
\delta_1 = \frac{1}{2}\pmatrix{-1\cr1\cr-1\cr1}, 
\delta_2 = \frac{1}{2}\pmatrix{-1\cr1\cr1\cr-1},
\delta_3 = \frac{1}{2}\pmatrix{1\cr1\cr1\cr1}, 
\delta_4 = \frac{1}{2}\pmatrix{1\cr1\cr-1\cr-1}.      \label{d}
\eeq
Proceeding in the same way as before one can derive the following equation for
the present representation:
\bea
P |\pm \delta_i \R = \pm \delta_i~ |\pm \delta_i \R~. 
\label{mon.deltai}
\eea
Again the following relations are true:
\bea
\begin{array}{ll}
|-\delta_1\R = E^{-\alpha_{11}}_0 |\delta_3\R~,&
|\delta_1\R = iE^{-\alpha_5}_0|\delta_3\R~,\\
|-\delta_2\R = -iE^{-\alpha_9}_0 |\delta_3\R~,&
|\delta_2\R = E^{-\alpha_7}_0 |\delta_3\R~, \\
|-\delta_3\R = iE^{-\alpha_1}_0 E^{-\alpha_3}_0 |\delta_3\R~, &\\  
|-\delta_4\R = -iE^{-\alpha_1}_0 |\delta_3\R~,&
|\delta_4\R = E^{-\alpha_3}_0|\delta_3\R~.\\
\end{array}
\label{pd3.rep.} 
\eea
\bea
\begin{array}{ll}
|\delta_1\R = -E^{\alpha_{11}}_0|-\delta_3\R~,&
|-\delta_1\R = iE^{\alpha_5}_0 |-\delta_3\R~,\\
|\delta_2\R = -iE^{\alpha_9}_0|-\delta_3\R~,&
|-\delta_2\R = -E^{\alpha_7}_0 |-\delta_3\R, \\
|\delta_3\R = -i E^{\alpha_1}_0 E^{\alpha_3}_0 |-\delta_3\R~, &\\
|\delta_4\R = -iE^{\alpha_1}_0|-\delta_3\R~,&
|-\delta_4\R = -E^{\alpha_3}_0 |-\delta_3\R~.\\
\end{array}
\label{md3.rep.}
\eea
\end{enumerate}

\sectiono{$\Gamma$-Matrix Representation}
\label{app-gamma}
\setcounter{equation}{0}
Let $\Gamma^{\mu}$ $(\mu =1,2,...,8)$ be the 16-dimensional $SO(8)$ Dirac 
matrices satisfying the Clifford algebra:
\bea
\lt\{\Gamma^{\mu}, \Gamma^{\nu} \rt\} = 2 \delta^{\mu \nu} {\bf 1}_{16}~. 
\label{Clif.alg.1}
\eea
${\bf 1}_{16}$ = $(16 \times 16)$ identity matrix. In the real-Weyl basis they 
take the following block-off-diagonal form:
\bea
\Gamma^{\mu} = \pmatrix{0 & \gamma^{\mu} \cr \bar \gamma^{\mu} & 0}~,
\label{box.off.gamma}
\eea
where $\gamma^{\mu}$ are $(8 \times 8)$ real matrices with $\bar \gamma^{\mu}
= \lt( \gamma^{\mu} \rt)^T$. In terms of $\gamma^{\mu}$'s, the algebra 
(\ref{Clif.alg.1}) reads:
\bea
\gamma^{\mu} \bar \gamma^{\nu} + \gamma^{\nu} \bar \gamma^{\mu} =
\bar \gamma^{\mu} \gamma^{\nu} + \bar \gamma^{\nu} \gamma^{\mu} =
2 \delta^{\mu \nu} {\bf 1}_8~.   \label{Clif.alg.2}
\eea 
${\bf 1}_8$ = $(8 \times 8)$ identity matrix. We take the following 
representation of the $\gamma$ matrices \cite{gsw}.
\bea
\begin{array}{ll}
\gamma^1= {\bf 1}  \times {\bf 1}  \times {\bf 1}~,  & \quad  
\gamma^2= \epsilon \times {\bf 1}  \times \sigma_3~, \\
\gamma^3= \sigma_3 \times \epsilon \times {\bf 1}~,  & \quad
\gamma^4= \sigma_1 \times \epsilon \times {\bf 1}~,  \\ 
\gamma^5= {\bf 1}  \times \sigma_3 \times \epsilon ~,& \quad
\gamma^6= \epsilon \times {\bf 1}  \times \sigma_1 ~,\\
\gamma^7= \epsilon \times \epsilon \times \epsilon ~,& \quad
\gamma^8= {\bf 1}  \times \sigma_1 \times \epsilon ~,
\end{array}  \label{gamma.matrices}
\eea
where ${\bf 1}$ = $(2 \times 2)$ identity matrix, $\epsilon = i\sigma_2$ and
$\sigma_1,~\sigma_2,~\sigma_3$ are the Pauli matrices:
\bea
\sigma_1 = \pmatrix{0&1\cr 1&0},~~ \sigma_2 = \pmatrix{0&-i\cr i&0},~~
\sigma_3 = \pmatrix{1&0\cr 0&-1}~.
\eea

\sectiono{Norm of Bosonic Vacuum states}
\label{app-basis}
\setcounter{equation}{0}
Here we will show that the norm of the bosonic vacuum states $|\vec s
\in {\cal K}^{(e)}\R$, defined in eqn.~(\ref{def.bos.vac1}), with respect to
hermitian inner product is $1$. From eqn.~(\ref{def.bos.vac1}) we can write,
\bea
\bigg| |\vec s \R \bigg|^2 &\equiv &\L \vec s | \vec s \R~, \nonumber\\
&=& \L V| \prod_i \lt\{ {\cal E}^{\alpha_i}(s_i)\rt\} ^{\dagger} 
\prod_j {\cal E}^{\alpha_j}(s_j) |V\R~, \nonumber\\
\cr 
&=& \L V| \lt\{ {\cal E}^{\alpha_4}(s_4)\rt\} ^{\dagger} \cdots
\lt\{ {\cal E}^{\alpha_1}(s_1\mp 1)\rt\} ^{\dagger}
\lt(E^{\mp \alpha_1}_{2|s_1|-1}~E^{\pm \alpha_1}_{-2|s_1|+1} \rt) \cr
&& {\cal E}^{\alpha_1}(s_1\mp 1) \cdots {\cal E}^{\alpha_4}(s_4)|V\R~, 
\eea
where the upper and lower signs of $\pm$ or $\mp$ used above corresponds to 
the 
fact that $s_1>0$ and $s_1<0$ respectively. The same notation will be followed
throughout the rest of the derivation. Now using the commutator between 
$E^{\mp \alpha_1}_{2|s_1|-1}$ and $E^{\pm \alpha_1}_{-2|s_1|+1}$
(see eqn.~(\ref{EE})) we can write:
\bea
\bigg| |\vec s \R \bigg|^2 &=& I_1 + I_2~, \label{I1+I2}
\eea
where,
\bea
I_1 &=& \L s_1 \mp 1, s_2,s_3 , s_4 | \lt( \mp \alpha_1.H_0 + 2|s_1|-1 \rt)
|s_1\mp 1, s_2, s_3, s_4 \R~, \nonumber\\
I_2 &=& \bigg|E^{\mp \alpha_1}_{2|s_1|-1} |s_1\mp 1,s_2, s_3, s_4 \R \bigg|^2~.
\eea 
For $I_1$, one uses the orthogonality of $\alpha_i$'s and the fact that
$\alpha_i^2=2$ to get,
\bea
I_1 &=& \lt( 1 + 2|s_1| \mp 2s_1\rt) \bigg| |s_1\mp 1,s_2,s_3,s_4 \R \bigg|^2~,
 \cr
&=& \bigg| |s_1\mp 1,s_2,s_3,s_4 \R \bigg|^2~,  
\label{I1}
\eea
since according to the rule mentioned above, $\mp 2s_1 = -2|s_1|$. 

For $I_2$, one may proceed as follows:
\bea
&&E^{\mp \alpha_1}_{2|s_1|-1} |s_1\mp 1,s_2, s_3, s_4 \R \nonumber\\
\cr
&&=E^{\mp \alpha_1}_{2|s_1|-1} E^{\pm \alpha_1}_{-2|s_1|+3}
|s_1\mp 2,s_2, s_3, s_4 \R~, \nonumber\\
\cr
&&=\mp \alpha_1.H_2 |s_1\mp 2,s_2, s_3, s_4 \R + 
E^{\pm \alpha_1}_{-2|s_1|+3} E^{\mp \alpha_1}_{2|s_1|-1}
|s_1\mp 2,s_2, s_3, s_4 \R~, \cr
&&~~~~~~~~~~~~~~~~~~~~~~~~~~~~~~~~~~~~~~~~~~~~~~~~~~~~~~~~~
\lt[ \hbox{see eqn.}~(\ref{EE}) \rt]\nonumber\\
\cr
&&= 0 + E^{\pm \alpha_1}_{-2|s_1|+3} E^{\mp \alpha_1}_{2|s_1|-1}
E^{\pm \alpha_1}_{-2|s_1|+5} |s_1\mp 3,s_2, s_3, s_4 \R~, \cr 
&&~~~~~~~~~~~~~~~~~~~~~~~~~~~~~~~~~~~~~~~~~~~~~~~~~~~~~~~~~
\lt[ \hbox{see eqn.}~(\ref{bos.vac.state}) \rt]\nonumber\\
\cr
&&= E^{\pm \alpha_1}_{-2|s_1|+3} \lt(\mp \alpha_1.H_4 \rt)
|s_1\mp 3,s_2, s_3, s_4 \R + \cr
&&~~~ E^{\pm \alpha_1}_{-2|s_1|+3} E^{\pm \alpha_1}_{-2|s_1|+5}
E^{\mp \alpha_1}_{2|s_1|-1} |s_1\mp 3,s_2, s_3, s_4 \R~ \nonumber\\
\cr
&&= 0 + E^{\pm \alpha_1}_{-2|s_1|+3} E^{\pm \alpha_1}_{-2|s_1|+5}
E^{\mp \alpha_1}_{2|s_1|-1} |s_1\mp 3,s_2, s_3, s_4 \R~. \cr
&&~~~~~~~~~~~~~~~~~~~~~~~~~~~~~~~~~~~~~~~~~~~~~~\,
\lt[ \hbox{again using eqn.}~(\ref{bos.vac.state}) \rt] \nonumber
\eea
Proceeding in the same way one finally gets,
\bea
E^{\mp \alpha_1}_{2|s_1|-1} |s_1\mp 1,s_2, s_3, s_4 \R &=&
{\cal E}^{\alpha_1}(s_1\mp 1)~ E^{\mp \alpha_1}_{2|s_1|-1} \prod_{i=2}^4
{\cal E}^{\alpha_i}(s_i) |V\R~. 
\eea
But since all $E^{\alpha_i}_n$'s commute, $E^{\mp \alpha_1}_{2|s_1|-1}$ in the 
above equation passes through all the ${\cal E}^{\alpha_i}(s_i)$'s, hits the
vacuum state $|V\R$ and annihilates it (as $2|s_1|-1 >0$). Therefore,
\bea
I_2 = 0~. \label{I2}
\eea
Therefore, using eqns. (\ref{I1+I2}), (\ref{I1}) and (\ref{I2}) we get,
\bea
\bigg||\vec s \R \bigg|^2 = \bigg| |s_1\mp 1,s_2,s_3,s_4 \R \bigg|^2~.
\eea
Repeating the above procedure to remove all the $E$ oscilltors sandwiched 
between $\L V|$ and $|V\R$, one can finally show:
\bea
\bigg| |\vec s \R \bigg|^2 = \bigg||V\R \bigg|^2 = 1~.
\label{normalvac}
\eea 

\section*{Acknowledgements}
I am extremely grateful to Ashoke Sen for suggesting the problem. His guidance 
and enormous help throughout the process of solving the problem and writing the
 paper has made this work possible. I would like to thank Debashis Ghoshal and 
Dileep P. Jatkar for their patient and critical reading of the preliminary 
draft.

\end{document}